\documentclass[a4paper,11pt]{article}

\pdfoutput=1
\usepackage{graphicx} 
\usepackage{jcappub} 

\usepackage[T1]{fontenc} 

\usepackage{graphicx} 
\graphicspath{{Figures/}}
\usepackage[utf8]{inputenc}
\usepackage{dcolumn} 
\usepackage{bm}      
\usepackage[colorlinks=true,linkcolor=blue,citecolor=blue]{hyperref}
\usepackage{verbatim}

\usepackage{xcolor}

\usepackage{braket}

\usepackage{siunitx}

\usepackage{xfrac,faktor}

\usepackage[normalem]{ulem}

\usepackage[capitalise]{cleveref}
\Crefname{section}{Sec.}{Sections~}

\usepackage{verbatim}

\usepackage{multirow}

\usepackage{bbold}

\DeclareUnicodeCharacter{2212}{-}
\DeclareUnicodeCharacter{2265}{>}

\usepackage{grffile}
\usepackage{amssymb,amsmath,graphicx,color,microtype}
\usepackage{hyperref}
\usepackage{verbatim}
\usepackage{enumerate}
\usepackage{subfigure}
\usepackage{multirow}
\usepackage{appendix}
\usepackage{setspace}
\usepackage{placeins}
\usepackage{booktabs}
\graphicspath{{figures/}}

\def \deltag{{\delta_g}}

\def \ba{\begin{eqnarray}}
\def \ea{\end{eqnarray}}
\def \be{\begin{equation}}
\def \ee{\end{equation}}

\def\pl{\parallel}

\def\cpl{\chi_\pl}
\def\cpli{\chi_{\pl,i}}

\def\qpar{q_{\pl}}

\def\dt21{\delta T_{21}}
\def\dtone21{\delta T_{21}^{(1)}}
\def\dttwo21{\delta T_{21}^{(2)}}
\def\dttwob21{\overline {\delta T_{21}^{(2)}}}
\def\dtonecc21{\delta T_{21}^{(1)*}}
\def\dttwobcc21{\overline {\delta T_{21}^{(2)*}}}

\def\bfk{{\bf k}}	
\def\bfq{{\bf q}}	
	
\def\bfr{{\bf r}}	
\def\dw21{{\delta^W_{21}}}
\def\d21{{\delta_{21}}}

\def\bfq{{\textbf{q}}}

\DeclareMathOperator{\sinc}{sinc}

\title{\boldmath Cross-Bispectrum Constraints from HI Intensity Mapping and Galaxy Surveys}

\author[a,b]{Warren Naidoo}
\author[a,b,c]{Moumita Aich}
\author[a,b]{Kavilan Moodley}

\affiliation[a]{Astrophysics Research Centre, University of KwaZulu-Natal, Westville Campus, Durban 4041, South Africa}
\affiliation[b]{School of Agriculture and Science, University of KwaZulu-Natal, Westville Campus, Durban 4041, South Africa}
\affiliation[c]{Wits Anglo American Digital Dome \& Wits Centre for Astrophysics, School of Physics, University of Witwatersrand, Johannesburg, South Africa}

\emailAdd{naidoow1@ukzn.ac.za} 

\date{June 2026}

\abstract{Future neutral hydrogen (HI) intensity mapping (IM) experiments and optical surveys will probe the large-scale structure of the universe over unprecedented volumes. Combining these complementary probes provides a promising avenue for constraining galaxy and cosmological models. In this work, we study the cross-correlation between the HI IM field and the galaxy density and cosmic shear fields probed by galaxy surveys. We formulate the HI-HI-galaxy and HI-HI-shear cross-bispectra, which provide sensitivity to long-wavelength radial modes otherwise removed by foreground subtraction in HI maps, and evaluate their detectability using measurements from the HIRAX telescope and the Rubin-LSST survey.

We find that combining the cross-bispectra with HI and galaxy or shear power spectra breaks degeneracies in redshift-dependent quantities, including the $f$-$\sigma_8$ degeneracy. Our constraints range from $0.74\%$ to $9.8\%$ on $f$ and $0.073\%$ to $3.5\%$ on $\sigma_8$ across the six redshift bins that we consider. Our constraints on the first- and second-order HI bias parameters range from $0.42\%$ to $5.6\%$ on $b_{\rm HI}^{(1)}$ and from $2.0\%$ to $29\%$ on $b_{\rm HI}^{(2)}$, while the galaxy bias constraints range from $2.6\%$ to $11\%$ on $b_{\rm gal}.$ The HI-HI-galaxy density cross-bispectrum measurement allows us to constrain a linear scale-dependent model for the stochastic cross-correlation coefficient between HI and stars, $r_{{\rm HI \,HI\,}\delta_g} = r_0 + r_1 k,$ with a break scale of $0.14\,\mbox{Mpc}^{-1}.$ We obtain $r_0$ constraints that range from $3.8\%$ to $13.5\%$ and $r_1$ constraints that range from $7.5\%$ to $19.9\%$. 

For cosmological constraints, we find that combining the cross-bispectra with HI and galaxy or shear power spectra provides additional constraining power on the $\Lambda$CDM model, particularly from the HI-HI-shear cross-bispectrum. For the $w_0\, w_a$CDM model, the power-spectrum-only constraints on $w_0$ and $w_a,$ including Planck priors, improve from $(\sigma_{w_0},\sigma_{w_a})=(0.036,0.15)$ and $(0.032,0.12)$ to $(\sigma_{w_0},\sigma_{w_a})=(0.014,0.057)$ and $(0.015,0.059)$ when adding the HI-HI-galaxy and HI-HI-shear cross-bispectra, respectively. }

\notoc

\begin{document}

\maketitle

\section{Introduction}

Intensity mapping (IM) of neutral hydrogen (HI) provides a promising approach for accurately measuring the baryon acoustic oscillations (BAO) signal \cite{battye2013h, Bull}. HI IM measures the large-scale structure of the universe via the 21cm line emission \cite{furlanetto2006cosmology}, serving as a biased tracer of the underlying dark matter distribution. IM experiments trace the large-scale structure using spectral lines, enabling surveys of larger cosmic volumes than traditional galaxy surveys. Several existing and forthcoming HI IM experiments, such as the Hydrogen Intensity and Real-time Analysis eXperiment (HIRAX) \cite{crichton2022hydrogen}, the Square Kilometer Array - Mid (SKA-Mid) \cite{ska}, Tianlai \cite{chen2012tianlai}, the Baryon Acoustic Oscillations from Integrated Neutral Gas Observations (BINGO) \cite{battye2016update}, the Canadian Hydrogen Intensity Mapping Experiment (CHIME) \cite{bandura2014canadian}, the Canadian Hydrogen Observatory and Radio-transient Detector (CHORD) \cite{vanderlinde2019lrp}, and the Five-hundred-meter Aperture Spherical radio Telescope (FAST/Tianyan) \cite{NAN_2011, Hu_2020}, aim to measure the HI distribution from low to intermediate redshifts. In this paper, we focus on the HIRAX intensity mapping survey.

In addition to HI surveys, optical surveys such as the Vera C. Rubin Observatory Legacy Survey of Space and Time (Rubin-LSST) \cite{lsst}, Dark Energy Spectroscopic Instrument (DESI) \cite{desi}, EUCLID \cite{euclid}, the Dark Energy Survey (DES) \cite{dark2016dark}, the Kilo Degree Survey (KIDS) \cite{de2013kilo}, the Hyper Suprime-Cam Subaru Strategic Program (HSC-SSP) \cite{aihara2018hyper} and the Nancy Grace Roman Space Telescope \cite{wfirst} will probe large-scale structure by mapping galaxy distributions across cosmic time. In this paper, we focus on the Rubin-LSST survey, relying on two key probes: cosmic shear, which probes the growth of structure and the angular diameter distance through weak gravitational lensing \cite{Bartelmann_2001}, and galaxy density measurements, which probe the dark matter power spectrum using galaxies as a biased tracer  \cite{lsstde, Ivanov_2020}. 

HI and optical surveys trace the same underlying large-scale structure, and previous work has studied the constraints resulting from the cross-correlation between these probes \cite{wyithe2007correlation, Chang:2010jp, masui2013measurement,  wolz2017determining, Pourtsidou, Alonso, padmanabhan2020cross}. 
Measurements of the cross-correlation power spectrum constrain the growth of large-scale structure and the relationship between HI gas and galaxy populations \cite{Pourtsidou, Alonso, fonseca2018synergies, padmanabhan2020cross}. They also provide a calibration of photometric redshifts \cite{Alonso, guandalin2022clustering}, probe lensing magnification \cite{jalilvand2020new}, constrain the HI fraction and bias, which remain poorly constrained at intermediate and high redshifts \cite{pourtsidou2017h, bacon2020cosmology}, and probe the connection between baryonic gas and star formation \cite{wolz2016intensity, guo2017constraining}. Furthermore, HI and galaxy cross-correlations provide a powerful tool for precision cosmology, probing extensions to the standard model \cite{Pourtsidou2016, Pourtsidou2015, ansari2018inflation, shi2020hir4, viljoen2021multi, fang2022cosmology, sgier2021combined, white2022cosmological}.

Astrophysical foreground contamination remains a major challenge for HI surveys, even in cross-correlation studies. Galactic synchrotron emission and extragalactic point sources are several orders of magnitude brighter than the HI signal \cite{Santos_2005}. Foreground mitigation that exploits the spectral smoothness of foreground emission \cite{Shaw_2014, Liu2013} removes large-scale radial HI modes or leaks foreground power into the recovered HI signal \cite{Shaw_2015, Switzer_2014}. The avoidance of the foreground wedge provides a cleaner measurement of the HI signal \cite{Parsons_2012} but significantly reduces the accessible HI Fourier modes. Detecting the HI signal in cross-correlation with cosmic shear or photometric galaxy surveys remains a challenge due to the loss of large-scale radial HI modes \cite{moodley2023crossbispectrum}. An alternative approach is to study wide-angle correlations, taking advantage of mode coupling from light-cone evolution to recover the large-scale cross-correlation \cite{kothari2024wide, shen2026direct}. 

Due to these challenges, the first detection of the HI signal was achieved in cross-correlation, combining optical data from the DEEP2 survey with HI observations from the Green Bank Telescope \cite{Chang:2010jp}. Subsequent detections have extended this approach to different redshift ranges and survey combinations \cite{masui2013measurement, anderson2018low}. More recently, a detection of the cosmological HI signal in cross-correlation with the eBOSS survey was reported by CHIME \cite{Amiri_2023}, along with a detection in cross-correlation of MeerKAT HI measurements with the WiggleZ Dark Energy Survey \cite{cunnington2023h}.
A direct detection of the HI auto power spectrum on small scales $k \gtrsim 1 $Mpc$^{-1}$ has been reported by MeerKAT at redshifts $\sim 0.3-0.4$ \cite{paul2026direct}, and by CHIME on scales of $0.27\mathrm{Mpc}^{-1}  \lesssim k \lesssim 1.01 \mathrm{Mpc}^{-1} $ at redshift $z\sim 1$.

In this paper, we propose using cosmic shear and galaxy density surveys to measure the long-wavelength radial modes that are removed by foreground cleaning of HI IM, following the approach outlined in \cite{moodley2023crossbispectrum}. This is achieved through a cross-bispectrum involving two small-scale HI modes and a large-scale galaxy density or cosmic shear radial mode. This technique exploits the density modulation effect, in which long-wavelength density fluctuations modulate small-scale HI fluctuations \cite{Bernardeau_2002, chiang2014, chiang2015, Takada2013}, inducing a correlation between long- and short-wavelength modes. This approach thus enables the measurement of the cross-correlation between HI IM and projected tracers, such as cosmic shear or photometric galaxies, through the cross-bispectrum. Such a measurement provides additional cosmological information that breaks degeneracies between cosmological parameters \cite{Takada_2003}. A recent attempt to measure the squeezed bispectrum between the HI signal from CHIME and the CMB lensing from Planck did not yield a detection \cite{Chakraborty_2026}.

The remainder of this paper is organised as follows. In Section \ref{sec:Cosmic Tracers} and Section \ref{sec:gal_and_shear}, we introduce the HI, galaxy density and cosmic shear fields. In Section \ref{sec:Cross-corr}, we demonstrate that the HI cross-power spectra with cosmic shear or galaxy density vanish, before introducing the cross-bispectrum estimator. In Section \ref{sec:Forecasts}, we present parameter constraints from the combined probes using the Fisher analysis approach. Finally, we conclude with a discussion in Section \ref{sec:conclusion}. In this work, we adopt the Planck 2018 cosmology and priors \cite{Aghanim}: $h=0.67$, $\Omega_M = 0.315$, $\Omega_\Lambda = 0.684$, $\Omega_k = 0.0$, $n_s = 0.965$, $\sigma_8 =  0.811$, $w_0 = -1.03$ and $N_{eff} = 2.99.$ All distances and scales are expressed in physical (Mpc), rather than $h^{-1}$Mpc, units.

\section{The Cosmological HI Signal}
\label{sec:Cosmic Tracers}
We now outline the theoretical framework used to model the 21cm signal from HI which at the redshifts we consider is an emission. The brightness temperature of the 21cm signal, $T_{21}$, can be decomposed into homogeneous and fluctuating contributions $T_{21} = \bar{T}(1+\delta_\mathrm{HI})$, where $\delta_\mathrm{HI}$ is the HI density contrast and $\bar{T}$ is the mean brightness temperature \cite{Bull}. 
The brightness temperature fluctuation is then written as
\begin{equation}
\delta T_{21}(\mathbf{k}; z) = \bar{T}(z) \delta_{\rm HI} = \bar{T}(z) \left( b_\mathrm{HI}(z)+ f(z) \mu^2_k \right) \delta_m(\mathbf{k}; z) 
\end{equation}
where $\delta_m$ denotes the matter density perturbation, $b_{\rm HI}(z)$ is the HI bias and $f(z)$ is the linear growth function. We include the contribution from redshift-space distortions, arising from the peculiar velocities of galaxies hosting HI, through the Kaiser factor, $f \mu_k^2$, where $\mu_k=k_\parallel/k$. The mean brightness temperature is given by \cite{Bull}
\begin{equation}
\bar{T}(z)=566h \left(H_0\over H(z) \right) \left(\Omega_{\rm HI}(z)\over 0.003\right)(1+z)^2 \text{\space  }\mu K.
\end{equation}
We adopt the parametrisation of the HI density parameter $\Omega_\mathrm{HI}$ from \cite{Bull}, and a parametric model for the HI bias from \cite{penin}. The redshift dependence of the matter density contrast is encoded in the linear growth factor, $D(z)$, such that $\delta_m(\mathbf{k}; z) = D(z) \delta_m(\mathbf{k}; z=0) $. 

Using the flat-sky approximation, the brightness temperature fluctuation can be written in terms of projected Fourier coordinates, $(\ell, y)$, as 
\begin{equation} 
\begin{aligned}
\delta T_{21}(\mathbf{\ell},y;z_i) = \frac{\delta T_{21}(k_\perp=\ell/\chi_{i}, k_\parallel=y/r_{\nu,i}; z_i)}{(\chi_{i})^2 r_{\nu,i}}.
\label{eq:HI_brightness_temperature}
\end{aligned}
\end{equation}
The transverse comoving distance, which reduces to the line-of-sight comoving distance in a spatially flat universe, is given by $\chi(z)=\int^z_0 c dz/H(z)$ \cite{Hogg}. The radial comoving distance corresponding to the observed frequency interval is $r_{\parallel} =c (1+z)^2/H(z) (\tilde{\nu}_p - \tilde{\nu}_i)= r_\nu(z) (\tilde{\nu}_p - \tilde{\nu}_i)$, where $\tilde{\nu} = \nu/\nu_{21}$ \cite{Bull}.
In the above equation, $\chi^2 r_\nu$ defines the projection factor that relates the volume element in the comoving Fourier space to the projected Fourier coordinates. The brightness temperature field is evaluated within a redshift bin centred at $z_i$. In this approximation, redshift-dependent quantities are assumed to evolve sufficiently slowly across the bin that they may be approximated by their values at the central redshift, $z_i$. For our analysis, we adopt four logarithmically spaced redshift bins within the HIRAX redshift range, centred at $z_i = [0.81, 0.95, 1.27, 1.95]$ with bin edges corresponding to $[0.78, 0.85, 1.07, 1.52, 2.55]$. The corresponding normalized frequency bandwidth of each bin is $\Delta \tilde{\nu}_i$. We have verified that our forecasts are insensitive to both the adopted redshift binning and the narrow-bin approximation. The angular power spectrum of the HI field follows from Equation \ref{eq:HI_brightness_temperature} as
\begin{equation}
\begin{aligned}
C^{21}_\ell(y; z_i) = {P_{21}(\mathbf{k},z_i) \over (\chi_{i})^2  r_{\nu,i}}  =  \frac{\left[\bar{T}(z_i) \left( b_\mathrm{HI}+ f \mu^2_k \right)D(z_i)\right]^2}{(\chi_{i})^2  r_{\nu,i}} P_m(\mathbf{k}),
\label{eq:HI_pow_spec}
\end{aligned}
\end{equation} 
where $P_m(\mathbf{k}) = P_m(\mathbf{k}_\perp=\boldsymbol{\ell}/\chi, k_\parallel=y/r_{\nu,i})$ is the isotropic matter power spectrum and $P_{21}(\mathbf{k},z_i) $ is the HI power spectrum in Fourier space.

\begin{table}[b!]
	\centering
	\begin{tabular}{|c|c|}
		\hline
		Parameter &	HIRAX Specification\\
		\hline
		S$_{area}$[$\text{deg}^2$] & 15000 \\
		\hline
		T$_\text{obs}$[years] & 4 \\
		\hline
		Bandwidth[MHz] & 400 - 800 \\
		\hline
		T$_{\text{inst}}$[K]& 50 \\
		\hline
		N$_{\text{dish}}$& 1024 \\
		\hline
		D$_{\text{dish}}$[m]& 6 \\
		\hline
		D$_{\text{min}}$[m]& 6 \\
		\hline
		D$_{\text{max}}$[m]& 270\\
		\hline
	\end{tabular}
	\caption{\label{tab:experiments} Experimental specifications for the HIRAX survey.}
\end{table}

To assess the detectability of HI power spectrum we compare it to the interferometric angular noise power spectrum given by \cite{Bull}
\begin{equation}
C^{N,21}_\ell(y; z_i)  = \frac{T^2_\mathrm{sys} \lambda^4}{A^2_e \nu_{21} n(u=\ell/2\pi)} \,,
\end{equation}
where $T_\mathrm{sys}$ is the system temperature, $\lambda$ is the measured wavelength, $A_e$ is the effective collecting area and $n(u)$ is the baseline density function. We consider the HI signal as measured by HIRAX, a proposed 1024-element interferometric array currently under construction in the Karoo desert, in South Africa, which will survey approximately 15,000 square degrees of the southern sky. HIRAX will consist of 6m dual-polarized dishes operating over the 400-800 MHz frequency band, corresponding to the redshift range, $0.775<z<2.55$. 
The key experimental specifications of HIRAX are summarized in Table \ref{tab:experiments}.

\begin{figure}[t!]
    \centering
    \begin{subfigure}
        {\includegraphics[width=0.49\linewidth]{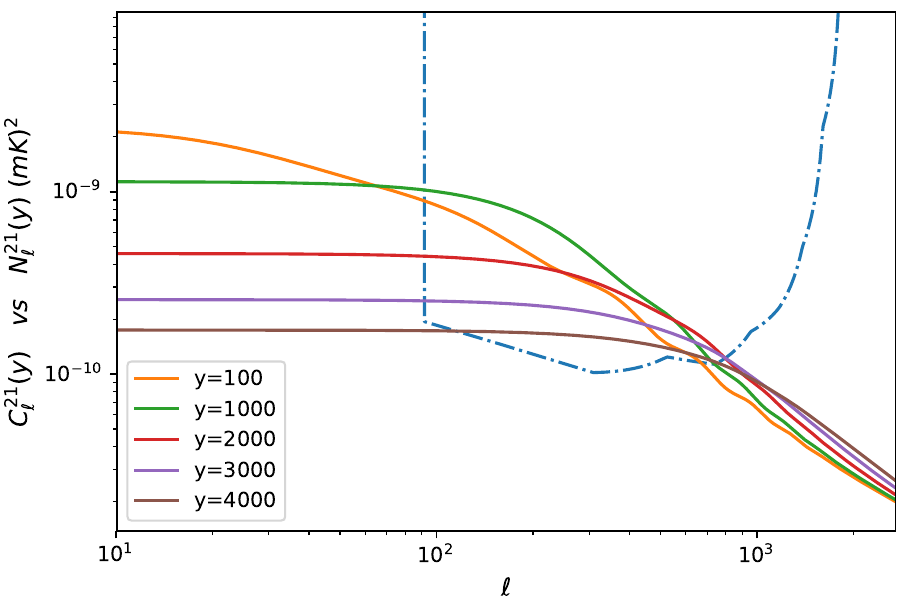}}
    \end{subfigure}
    \begin{subfigure}
        {\includegraphics[width=0.49\linewidth]{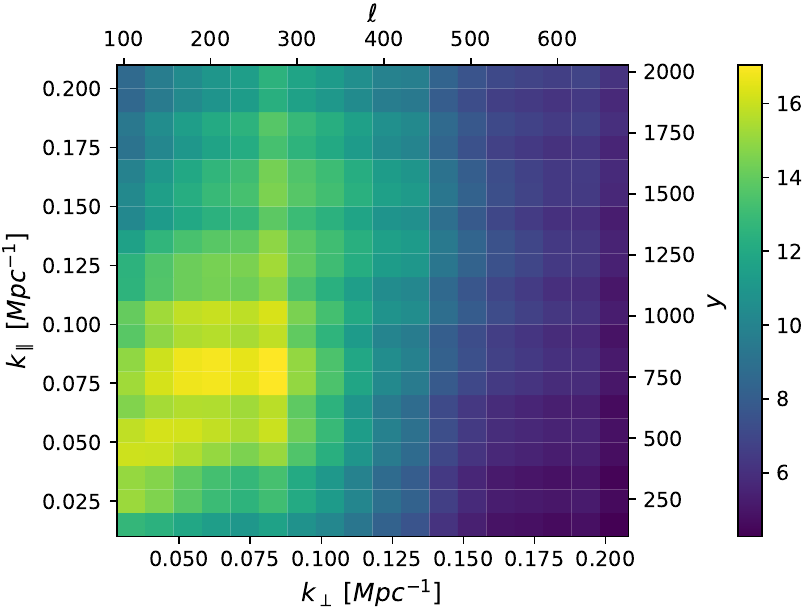}}
    \end{subfigure}
\caption{The HI cosmological signal vs the HIRAX instrument noise ({\it{left}}) computed at the $z_i=0.95$ redshift bin for an extended $\ell$ and $y$ range. We then show the signal-to-noise ratio ({\it{right}}) computed over the $k_\perp-k_\parallel$ plane in 0.01 Mpc$^{-1}$ bins for the same redshift restricted only to the linear regime.}
\label{fig:HI_sig_noise_snr}
\end{figure}

In Figure \ref{fig:HI_sig_noise_snr} ({\it{left}}), we show the HI signal and HIRAX noise power spectra as a function of $\ell$ for a set of radial, $y$, modes in a redshift bin centred at $z_i = 0.95$. 
Although we display an extended range of angular and radial modes, the modes included in our subsequent analysis are limited by instrumental and physical considerations. The accessible angular modes are set by the minimum ($D_{min}$) and maximum ($D_{max}$) baseline lengths of the interferometer, $\ell_{min} = 2\pi D_{min}/\lambda$ and $\ell_{max} = 2\pi D_{max}/\lambda$. Similarly, the radial modes are limited by the survey bandwidth and frequency channelization, $y_{min} \sim 1/\Delta \tilde{\nu}$ and  $y_{max} \sim 1/ \delta \tilde{\nu}$, where $\Delta \tilde{\nu} = \Delta  \nu / \nu_{21} $ and $\delta \tilde{\nu} = \delta  \nu / \nu_{21} $ are the normalized bandwidth and channel width, respectively. For the remainder of this work, we restrict our analysis to the linear regime, imposing $k < k_{NL}$ where $k_{NL} = 0.14(1+z)^{2/(2+n_s)}$.

Using the signal and noise power spectra of the HI field, we compute the signal-to-noise ratio in a given redshift bin, $z_i$, as
\begin{equation}
\begin{aligned}
\!\! (\mathrm{SNR}_i)^2 \!\! &= \! \frac{\Delta \tilde{\nu}_i S_\mathrm{area}}{2} \int^{y_\mathrm{max}}_{y_\mathrm{min}} \!\frac{dy}{2\pi} \int^{\ell_\mathrm{max}}_{\ell_\mathrm{min}} \frac{\ell d\ell}{2\pi} \frac{\mathcal{S}_i(\ell,y)^2}{\mathcal{V}_i(\ell,y)},
\label{SNRbin}
\end{aligned}
\end{equation}
where $\mathcal{S}_i(\ell,y)$ represents the signal of a given field and $\mathcal{V}_i(\ell,y)$ is its variance. For the projected HI power spectrum, we have $\mathcal{S}_i(\ell,y) = C^{21}_\ell(y; z_i)$ and $\mathcal{V}_i(\ell,y) =  (C^{21}_\ell(y; z_i) + C^{N,21}_\ell(y; z_i) )^2$. Accounting for the scale limits discussed above, we find signal-to-noise ratios of SNR$=[126, 206, 239, 192]$ across the four redshift bins respectively. The variation of the signal-to-noise ratio across redshift bins is partly driven by our logarithmic binning scheme, which distributes the available SNR more uniformly across redshift. 
In Figure \ref{fig:HI_sig_noise_snr} ({\it{right}}) we show the signal-to-noise ratio as a function of transverse and radial modes within the linear regime. We find that HIRAX provides strong detection prospects, particularly on BAO scales. 

For HI IM power spectrum detection, it is also important to account for scale cuts resulting from foreground contamination. Astrophysical foregrounds are typically several orders of magnitude larger than the cosmological HI signal. Since these foregrounds are generally smooth functions of frequency compared to the HI signal, their contamination can be mitigated by removing the largest-scale radial modes. For the above SNR results and the results in the sections that follow, we have applied a foreground cut of $k_\parallel < k_{FG}$ where $k_{FG} \sim 0.01 \text{ Mpc}^{-1}$ \cite{Bull}. 

\section{The Galaxy Density and Cosmic Shear Signals}
\label{sec:gal_and_shear}
Galaxy density and cosmic shear provide complementary probes of the underlying matter distribution, capturing information from both the biased tracer field and gravitational lensing of matter along the line of sight. The galaxy density and cosmic shear fields are given as \cite{Tegmark_2002, Peiris_2000, Kaiser_1998, Bartelmann_2001, Barber, Joachimi_2010, Kilbinger_2015,Kilbinger_2017}
\begin{equation}
    X(\ell) = \int \frac{dk_\parallel}{2\pi} \int d\chi W_{X}(\chi) \frac{e^{ik_\parallel \chi}}{\chi^2} \delta_m(k_\perp=\ell/\chi,k_\parallel)
\label{eq:gal_field}
\end{equation}
where $X = \deltag $ or $X = \gamma$ for galaxy density or cosmic shear respectively. The corresponding galaxy density and cosmic shear kernels are
\begin{equation}
W_{\deltag}(\chi) =  D(\chi) b_\mathrm{gal}(\chi) \frac{dN(z)}{dz} \frac{dz}{d\chi} 
\end{equation}
and
\begin{equation}
W_\gamma\left(\chi\right) = \frac{3}{2} \Omega_{m0} \left( \frac{H_0}{c}\right)^2 \chi \, g(\chi) D(\chi) (1+z).
\end{equation}
Here, $b_\mathrm{gal}$ is the galaxy bias which describes the clustering of galaxies relative to the underlying dark matter distribution. The galaxy distribution function is denoted by $dN/dz$ \cite{lsst, Alonso, fang, song}. We neglect the magnification bias, which accounts for a change in source density due to gravitational lensing, since it is subdominant to $dN/dz$ on the scales we consider \cite{Sherwin_2012}.
The lensing efficiency function $g(\chi)$ is given by
\begin{equation}
g(\chi) = \int_{\chi}^{\chi_\mathrm{lim}} d\chi' \frac{\chi'-\chi}{\chi'}  n(\chi'),
\end{equation}
where the integral extends to the limiting comoving distance $\chi_\mathrm{lim}$ of the source galaxy distribution which is described by $n(\chi) = \frac{dN}{dz} \frac{dz}{d\chi}$. 

Using the galaxy density and cosmic shear fields we compute their angular power spectra as
\ba 
C^{X}_\ell  &=& \frac{1}{2\pi} \int dk_\parallel \int d\chi W_{X} (\chi) \int d\chi' W_{X}(\chi')  \frac{e^{ik_\parallel(\chi-\chi')}}{\chi'^2} P_m(\mathbf{k}).
\label{gal-auto}
\ea 
To evaluate the power spectra, we use the Rubin-LSST 10-year survey `gold sample', for which the galaxy distribution function is ${dN/}{dz} = n_{\mathrm{gal}} \, p(z)$ \cite{mandelbaum2018lsst}, where the galaxy surface density is $ n_{\mathrm{gal}} = 172,800/\mbox{deg}^2,$ and
\begin{equation}
p(z) =  {  z^2  \exp(-(z/z_0)^\alpha) \over \int dz \, z^2  \exp(-(z/z_0)^\alpha)   } \, ,
\end{equation} 
with $z_0=0.28$ and $\alpha=0.9$  \cite{mandelbaum2018lsst}. The redshift distribution is normalised such that its integral over the full redshift range of the survey equals the galaxy surface density. The Rubin-LSST will conduct a photometric galaxy survey with a field of view of 9.6 deg$^2$, observing approximately 20,000 deg$^2$ of sky area in six broad photometric bands. We adopt the galaxy bias $b_\mathrm{gal}(z) = 1 + 0.84 z$ for the gold sample \cite{lsst}.

\begin{figure}[!t]
    \centering
    \begin{subfigure}
        {\includegraphics[width=0.49\linewidth]{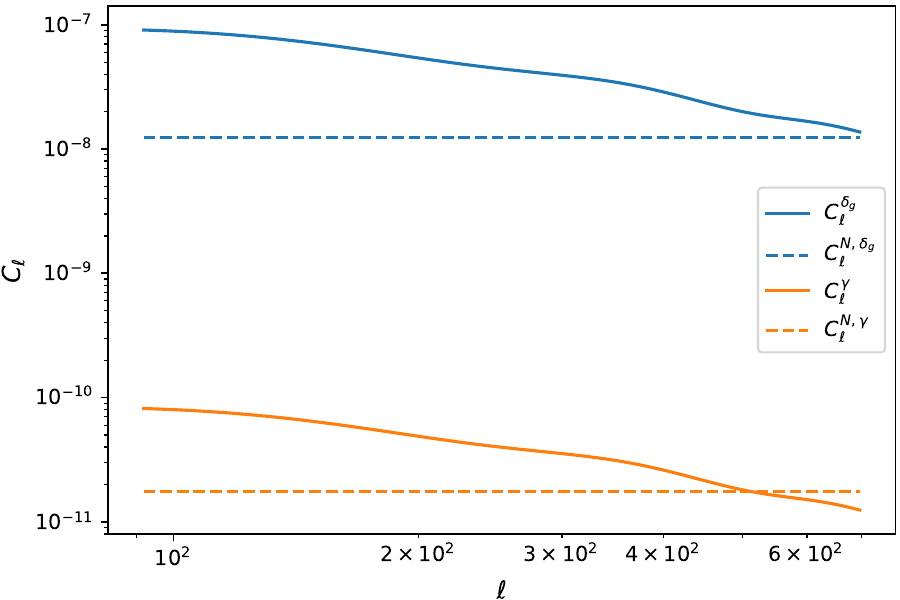}}
    \end{subfigure}
    \begin{subfigure}
        {\includegraphics[width=0.49\linewidth]{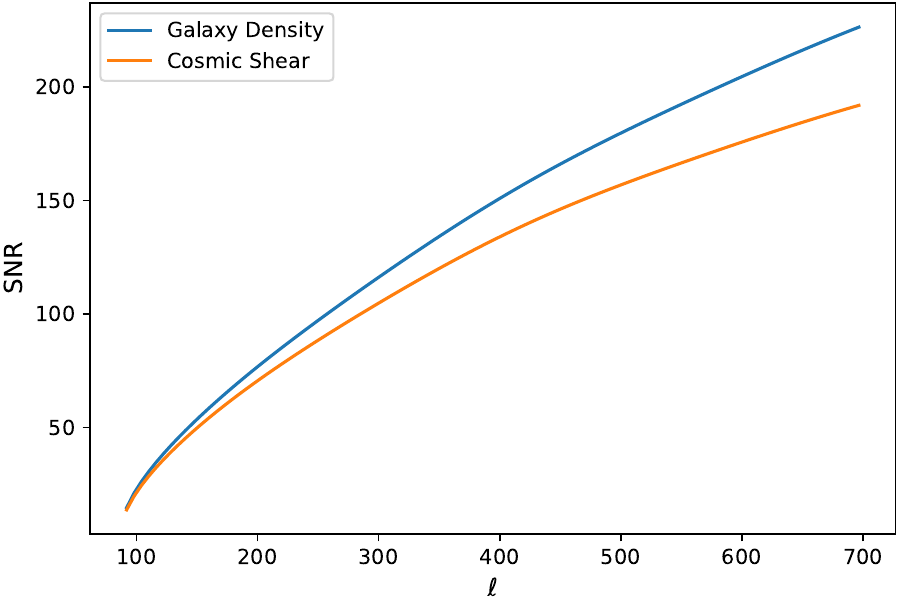}}
    \end{subfigure}
    \caption{The galaxy density and cosmic shear power spectra vs shot-noise power spectra for a single redshift bin centered at $z_i = 0.95$ ({\it{left}}). The corresponding cumulative signal-to-noise ratios as a function of angular wavenumber for the same redshift bin ({\it{right}}).}
    \label{fig:Gal_and_shear_auto_snr}
\end{figure}

We wish to compare the signal spectra of the galaxy density and cosmic shear surveys with their respective noise contributions from the Rubin-LSST survey. The power spectrum of shot noise for galaxy density surveys arises from the statistical uncertainty associated with discrete galaxy counts \cite{Tegmark_2002}. The cosmic shear noise power spectrum includes an additional uncertainty factor $\langle \gamma^2_\mathrm{int} \rangle ^{(1/2)}$, which describes the intrinsic RMS shear per galaxy due to intrinsic ellipticities \cite{Kaiser_1992, Hu_2001}. We therefore write the angular noise power spectra for the galaxy density and cosmic shear fields as
\ba 
C^{N,\mathrm{X}}_\ell  = \mathcal{C}_\mathrm{X} \frac{1}{\left(\int^{z_\mathrm{max}}_{z_\mathrm{min}}  dz \frac{d N(z)}{dz} \right) / V_\mathrm{sur}},
\ea
such that $ \mathcal{C}_\deltag = 1$ for galaxy density, and $\mathcal{C}_\mathrm{shear} = \langle \gamma^2_\mathrm{int} \rangle ^{(1/2)} = 0.8$ for the cosmic shear case corresponding to $\langle \gamma^2_\mathrm{int} \rangle ^{(1/2)}$. The total number of galaxies in a given redshift bin of the survey is given by $\int^{z_\mathrm{max}}_{z_\mathrm{min}}  dz \frac{dN(z)}{dz}$  and the corresponding survey volume $V_\mathrm{sur}$ is \cite{Pourtsidou2016}
\be
V_\mathrm{sur} = S_\mathrm{area} \int^{z_\mathrm{max}}_{z_\mathrm{min}} dz \frac{c \chi(z)^2}{H(z)} \,.
\ee
In Figure \ref{fig:Gal_and_shear_auto_snr} ({\it{left}}) we compare the angular power spectra with the shot-noise power spectra for galaxy density (solid curve) and cosmic shear (dashed curve). The shot noise contribution is subdominant to the galaxy density and cosmic shear power spectra on most angular scales. We also see that the cosmic shear field is a less sensitive probe than the galaxy density field, due to the higher surface density of source galaxies compared to the effective lensing density.

We use the galaxy density and cosmic shear signal power spectra and shot noise power spectra to compute the signal-to-noise ratios 
\ba 
(\mathrm{SNR}_i)^2  = \int d \ell \frac{(2\ell+1)  f_\mathrm{sky}}{2} \left(\frac{C^\mathrm{S}_\ell}{C^\mathrm{S}_\ell+C^\mathrm{N}_\ell}\right)^2 \,, 
\ea
where $f_\mathrm{sky}$ is the sky fraction of the survey. In Figure \ref{fig:Gal_and_shear_auto_snr} ({\it{right}}) we show the cumulative signal-to-noise ratio as a function of angular wavenumber for galaxy density (solid curve) and cosmic shear (dashed curve). The Rubin-LSST 10-year survey shows strong prospects for detecting both fields. 

\section{HI Cross-Correlations with Galaxy Density and Cosmic Shear}
\label{sec:Cross-corr}
Cross-correlation provides a powerful tool for probing large-scale structure by mitigating survey-specific systematics and uncorrelated noise. We consider the cross-correlation of the HIRAX HI survey with galaxy density and cosmic shear surveys from the Rubin-LSST observatory. We compute the cross-correlation in a comoving volume, $V_L(z_i)$, spanning a redshift bin of width $\Delta z$ centered at $z_i,$ and subtending a solid angle, $\Omega_i,$ on the sky. We first study the cross-power spectrum before considering the cross-bispectrum.

\subsection{The cross-power spectrum}
The two-point cross-correlation angular power spectrum between the HI and galaxy density or cosmic shear fields is given by
\begin{equation}
\begin{aligned}
C_\ell^{21,\mathrm{X}}(y; z_i) = r_{\mathrm{HI}-\mathrm{X}} \int \frac{d^2\ell'}{(2\pi)^2} \langle \delta T_{21}(\ell,y;z_i) \mathrm{X}^*(\ell') \rangle 
= \frac{\bar{T}(z_i) \left( b_\mathrm{HI}+ f \mu^2_k \right) D(z_i)K^{*}_\mathrm{X}(k_\parallel)}{V_p(z_i)}   P_m(\mathbf{k}), 
\label{21cmshear_cross}
\end{aligned}
\end{equation}
where `X' represents the galaxy density, $\deltag$, or cosmic shear, $\gamma$, fields. The final expression follows from substituting Equations \ref{eq:HI_brightness_temperature} and \ref{eq:gal_field} into the above integral. We define the radial Fourier space kernel as $K^{*}_\mathrm{X}(k_\parallel) = \int d \chi' e^{-i\chi k_\parallel} W_\mathrm{X}(\chi)D(\chi)$. The cross-correlation coefficient $r_{\mathrm{HI}-\mathrm{X}} = C_\ell^{21,\mathrm{X}} /\sqrt{C_\ell^{X} C_\ell^{21}}$ quantifies the degree of correlation between the two tracers of the dark matter distribution. For galaxy number counts, stochastic effects induce departures from perfect correlation, such that $r_{\mathrm{HI}-\deltag}\neq 1$. In contrast, the cosmic shear field is a direct line-of-sight projection of the matter density and not subject to stochastic biasing. We therefore assume $r_{\mathrm{HI}-\gamma}=1$. 

\begin{figure}[!t]
    \centering
    \begin{subfigure}
        {\includegraphics[width=0.49\linewidth]{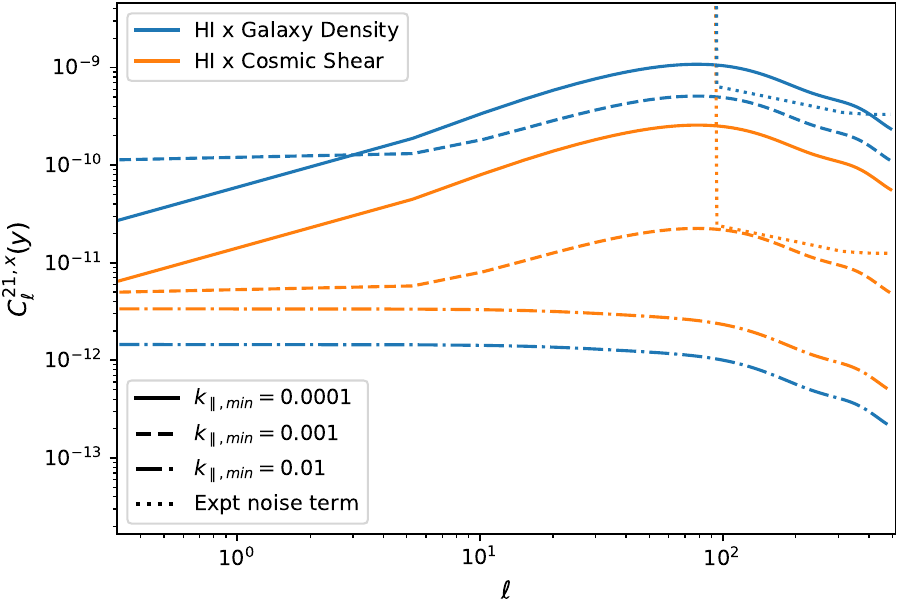}}
    \end{subfigure} 
    \begin{subfigure}
        {\includegraphics[width=0.49\linewidth]{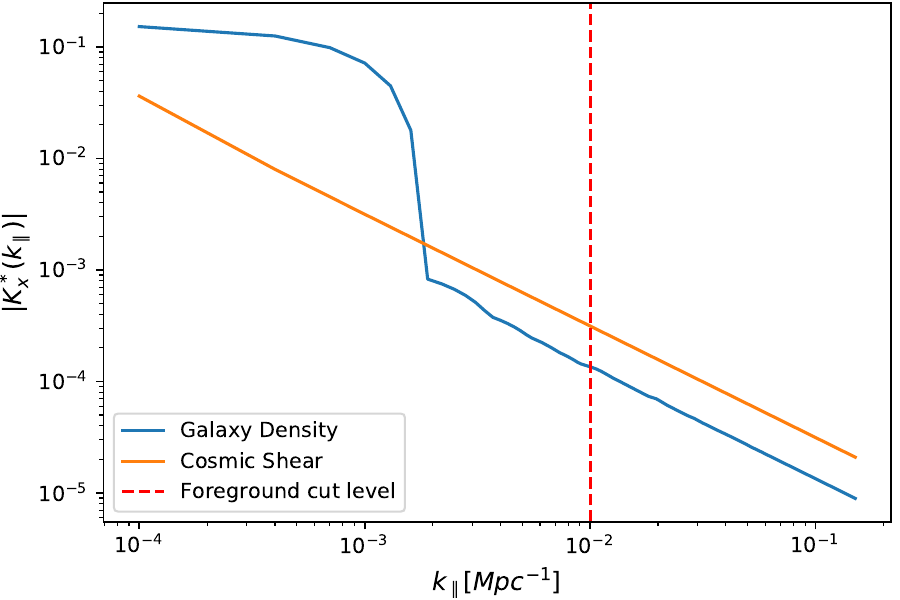}}
    \end{subfigure} 
    \caption{The HI-galaxy density/cosmic shear cross-correlation signals vs the experimental noise term ({\it{left}}) computed at the $z_i=0.95$ redshift bin. We show the effect of removing smoothly varying radial modes on the cross-correlation signal. The galaxy density and cosmic shear kernels in Fourier space, which the cross-correlation signal is directly proportional to, is also shown ({\it{right}}). We observe that the kernels decrease with increasing $k_\parallel$.}
    \label{fig:Gal_and_shear_2pt_cross}
\end{figure}

In Figure \ref{fig:Gal_and_shear_2pt_cross} ({\it{left}}) we show the cross-correlation power spectra for the HI and galaxy density (solid curve) or cosmic shear (dashed lines) as a function of $\ell$ for a few choices of radial mode cut, $k_{\parallel, min}$. We find that applying the nominal foreground cut, $k_{\parallel, min} = 0.01$ Mpc$^{-1}$, reduces the signal by several orders of magnitude, and well below the instrumental noise. This occurs because the galaxy density and cosmic shear measurements are primarily sensitive to the radial modes, which in the HI case are lost to foreground removal. Hence, following foreground removal, these cross-correlations do not produce a strongly detectable signal. 

To understand why the cross-power spectra vanish, we consider the Fourier space kernels for the galaxy density and cosmic shear fields shown in Figure \ref{fig:Gal_and_shear_2pt_cross} ({\it{right}}), which decrease in amplitude with increasing $k_\parallel$. The galaxy density kernel has enhanced support at $k_\parallel < 2\times 10^{-3}$ Mpc$^{-1}$ due to the support from narrower redshift bins, compared to the broader redshift bins necessary for the cosmic shear field. Nevertheless, the kernels for both fields have a significantly lower amplitude above $k_{\parallel,min}=0.01$ Mpc$^{-1}$, which significantly reduces their cross-power spectra. Since the cross-power spectra vanish, in the next subsection we consider the cross-bispectrum.

\subsection{The Cross-Bispectrum}
We now consider the HI cross-bispectra with galaxy density or cosmic shear, which relies on a density modulation effect \cite{Bernardeau_2002, chiang2014, chiang2015, Takada2013} to obtain a non-vanishing signal. The cross-bispectrum couples two small-scale HI modes to a long-wavelength galaxy density or cosmic shear mode \cite{moodley2023crossbispectrum}. We implement the cross-bispectrum using the integrated bispectrum approach \cite{chiang2014, chiang2015}. Since the bispectrum vanishes for Gaussian first-order fields, we consider the HI field to second order as \cite{Bernardeau_2002, moodley2023crossbispectrum}
\be
\delta T_{21}(\bfk; z_i) = \delta T^{(1)}_{21}(\bfk; z_i) + \delta T^{(2)}_{21}(\bfk; z_i)/\bar{T_b}(z_i) \,
\ee
to obtain a nonzero cross-correlation signal. We compute the cross-bispectrum signal over a sub-volume $V_L(z_i) = \Omega_i \Delta \tilde{\nu}_i V_p(z_i)$, corresponding to the dimensionless bandwidth across the redshift bin, subtending a solid angle, $\Omega_i $, on the sky. The HI field computed in the sub-volume is
\begin{equation}
\dt21(\bfk; z_i) \vert_\bfr
=V_{L} \int {d^3 k_1 \over (2\pi)^3} e^{-i\bfk_1 \cdot \bfr} \mathcal{W}_{L,21}(\bfk_1) \dt21\left (\bfk - \bfk_1; z_i \right),
\end{equation}
where $\mathcal{W}_{L,21}(\bfq') =  \mathcal{W}_{L,21}^\parallel(\qpar' L_\parallel) \times \, \mathcal{W}_{L,21}^\perp(q_\perp' L_\perp)$ is the window function over the sub-volume $V_L = L_\parallel L_\perp^2$. We use a top-hat window function in position space, corresponding to $\mathcal{W}_L(q) = \sinc(Lq)$ in Fourier space. To obtain the average galaxy density or average cosmic shear in the sub-volume we first define the average density field in a volume centered at position $r_i$ as \cite{chiang2014, chiang2015}
\begin{equation}
	\bar{\delta}(\mathbf{r}_i) = \int d^3 r \delta(\mathbf{r}) \mathcal{W}(\mathbf{r} - \mathbf{r}_i)
\end{equation}
where $\mathbf{r} = [\boldsymbol{r}_\perp = \chi \boldsymbol{\theta} , r_\parallel = \chi]$. Applying this to the galaxy density or cosmic shear fields and writing the matter density field in Fourier space we get \cite{moodley2023crossbispectrum}
\begin{equation}
\begin{aligned}
\bar{\mathrm{X}}(\bfr_i; z_i) &= {V_{L}  } \left[W_\mathrm{X}(\cpli) D(\cpli) /  \cpli^2 \right] \int {d^3 q' \over (2\pi)^3} \, \delta_m(-\bfq'; z=0) \, e^{-i \bfr_i \cdot \bfq'} \, \mathcal{W}_{L,\mathrm{X}}(\bfq'),
\end{aligned}
\end{equation}
where $\mathcal{W}_{L,\mathrm{X}}(\bfq') $ is the galaxy density or cosmic shear window function over the sub-volume.  The cross-bispectrum is then computed as follows \cite{moodley2023crossbispectrum}: \newpage
\begin{equation}
\begin{aligned}
    &B^{2121\mathrm{X}}_\ell (y; z_i)= \left\langle {1\over V_{L} V_p } \dt21(\bfk; z_i) \vert_\bfr \dt21^*(\bfk; z_i) \vert_\bfr  \bar{\mathrm{X}}^{*}\left(\bfr_i\right) \vert_\bfr \right\rangle   \\ &=  {1\over V_{L} V_p }  \left\langle\left[\dt21^{(1)}(\bfk; z_i) \vert_\bfr +\overline{\dt21}^{(2)}(\bfk; z_i) \vert_\bfr \right] \left[\dt21^{*(1)}(\bfk; z_i) \vert_\bfr +\overline{\dt21}^{*(2)}(\bfk; z_i) \vert_\bfr \right] \bar{\mathrm{X}}^{*}\left(\bfr_i\right)\vert_\bfr  \right\rangle \\
    \\ &=  {1\over V_{L} V_p }  \left\langle\left[\dt21^{(1)}(\bfk; z_i) \vert_\bfr \overline{\dt21}^{*(2)}(\bfk; z_i) \vert_\bfr + \dt21^{*(1)}(\bfk; z_i) \vert_\bfr \overline{\dt21}^{(2)}(\bfk; z_i) \vert_\bfr \right] \bar{\mathrm{X}}^{*}\left(\bfr_i\right) \vert_\bfr \right\rangle, \\
\end{aligned}
\end{equation}
where we have neglected fourth and higher order terms. Since the first-order fields are Gaussian, the bispectrum term $\langle \delta T_{21}^{(1)} \delta T_{21}^{(1)}\bar{\mathrm{X}}^{*} \rangle$ also vanishes.

\begin{figure}[!t]
\centering
\includegraphics[scale=0.6]{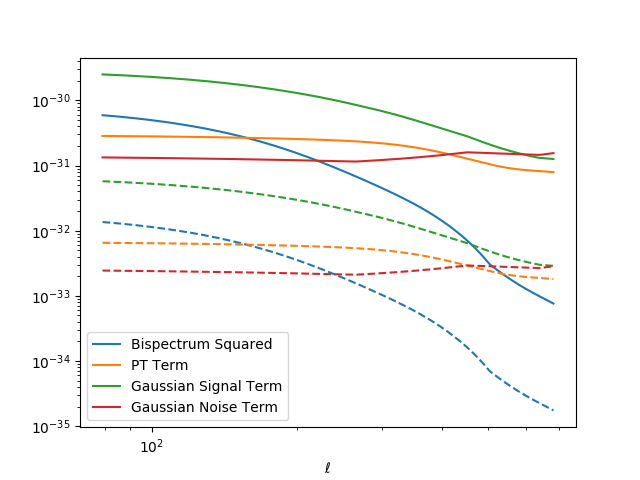}
\caption{The various cross-bispectrum variance terms evaluated at $y=917$ for the galaxy density (solid curves) and cosmic shear (dashed curves) showing the gaussian term dominates the variance contribution.}
\label{shear_var_terms}
\end{figure}
The final form of the cross-bispectrum is obtained by applying the squeezed-limit approximation, since the HI wavenumbers are much larger than the galaxy density or cosmic shear wavenumber. We compute the HI-HI-galaxy and HI-HI-shear cross-bispectra in a redshift bin centered at $z_i$, following \cite{moodley2023crossbispectrum} which computes the HI-CMB lensing cross-bispectrum, to obtain
\begin{equation} 
\begin{aligned}
B^{2121\mathrm{X}}_\ell (y; z_i) =
r_{\mathrm{HI\, HI\, X}} \frac{V^2_{L,j} W_\mathrm{X}(\chi_{i}) D(\chi_{i}) }{V_p(z_i) \chi^2_{i}}  
 P_{21}(\mathbf{k},z=0) \mathcal{B}(k,\mu_k; f, b_{\mathrm{HI}}^{(1)},b_{\mathrm{HI}}^{(2)})\\ \times \int q^2 d q \mathcal{W}_{L,\mathrm{X}}(q) \mathcal{W}_{L,21}(q) P_m(q;z=0),
\label{bispec1}
\end{aligned}
\end{equation}
where we have defined
\begin{equation}
\begin{aligned}
	\mathcal{B}(k, \mu_k, f, b_{\mathrm{HI}}^{(1)}, b_{\mathrm{HI}}^{(2)}) =  {1\over 3} \left( 3 - {d \log{P_m} \over d \log{k}} \right)  \left(f\mu_k^2 - \mu_k^2 + 2  \right)
    + {1 \over 14 \left(b_{\mathrm{HI}}^{(1)} + f\mu_k^2  \right) } \left(14b_{\mathrm{HI}}^{(1)}f\mu_k^2 \right. \\ \left. + {14\over 3}b_{\mathrm{HI}}^{(1)} f  + {26\over 3} b_{\mathrm{HI}}^{(1)} \mu_k^2 + {26 \over 3}b_{\mathrm{HI}}^{(1)} + 28b_{\mathrm{HI}}^{(2)} + 14f^2\mu_k^4 - 14f^2\mu_k^2 - 6f\mu_k^4 + {38\over 3}f\mu_k^2 \right).
\end{aligned}
\end{equation}
The cross-bispectrum, $B_{\ell}^{\mathrm{patch}},$ given in Equation \ref{bispec1} is computed over the patch of sky, $\Omega_i = 4\pi f_{\mathrm{patch}},$ corresponding to the chosen sub-volume. The full sky cross-bispectrum is given by $B_{\ell}^{\mathrm{sky}} = B_{\ell}^{\mathrm{patch}}/ f_{\mathrm{patch}}.$  

The bispectrum cross-correlation coefficient, $r_{\mathrm{HI\,HI\,X}} = B^{2121\mathrm{X}}_\ell (y; z_i)/\sqrt{\left(C_\ell^{21}(y;z_i)\right)^2 C_\ell^{X}},$ appearing in the bispectrum expression above, accounts for the stochasticity between the HI field and the galaxy density field.
In contrast, we set $r_{\mathrm{HI\,HI\,}\gamma} = 1$, as cosmic shear is not subject to stochastic bias relative to the HI field. We adopt the first and second-order HI biases,  $b_\mathrm{HI}^{(1)}$ and $b_\mathrm{HI}^{(2)}$, from \cite{penin}. From Equation \ref{bispec1}, it is evident that the cross-bispectrum can be interpreted as the cross-correlation between the HI position-dependent power spectrum and the average of a background tracer \cite{chiang2014, chiang2015}.

\begin{figure}[!t]
\includegraphics[width=0.49\linewidth]{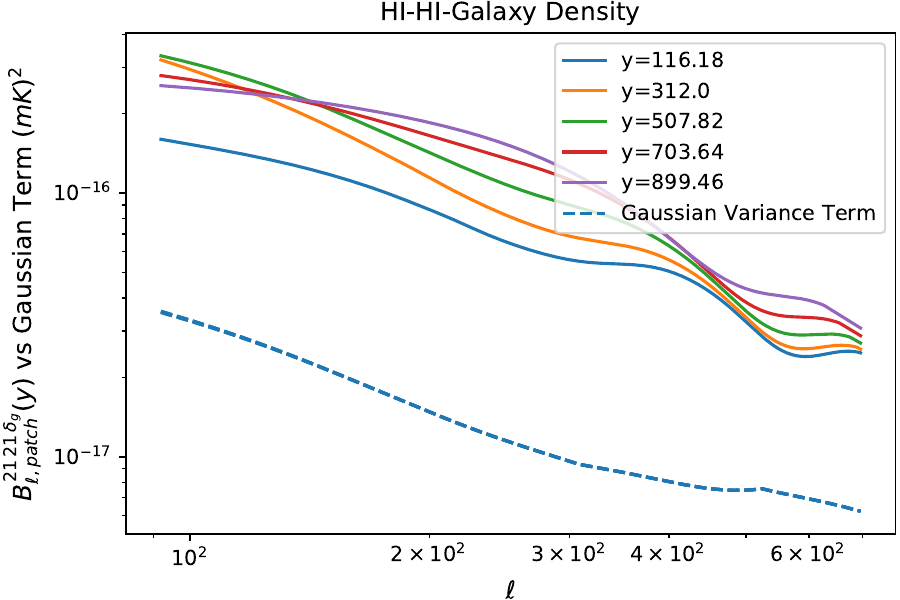}
\includegraphics[width=0.49\linewidth]{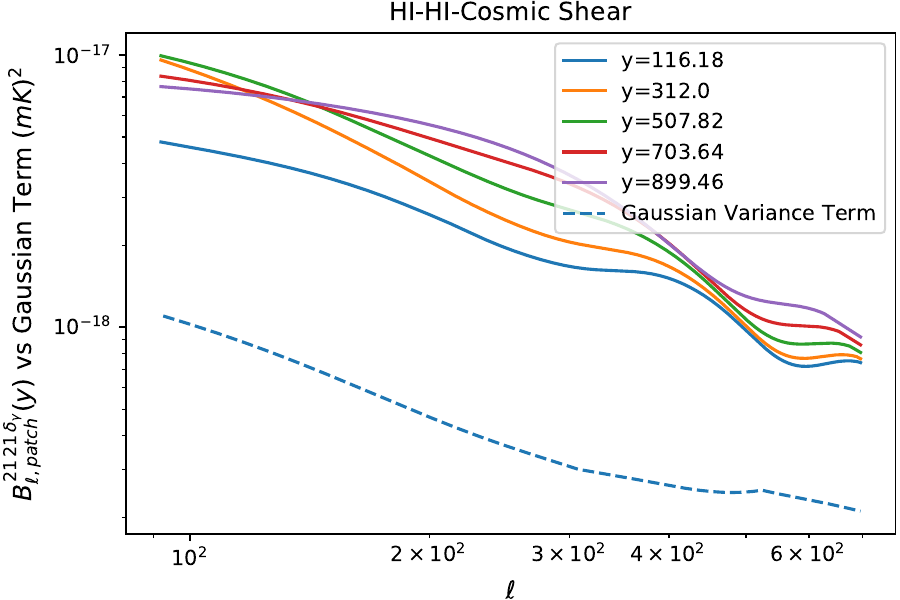}
\includegraphics[width=0.49\linewidth]{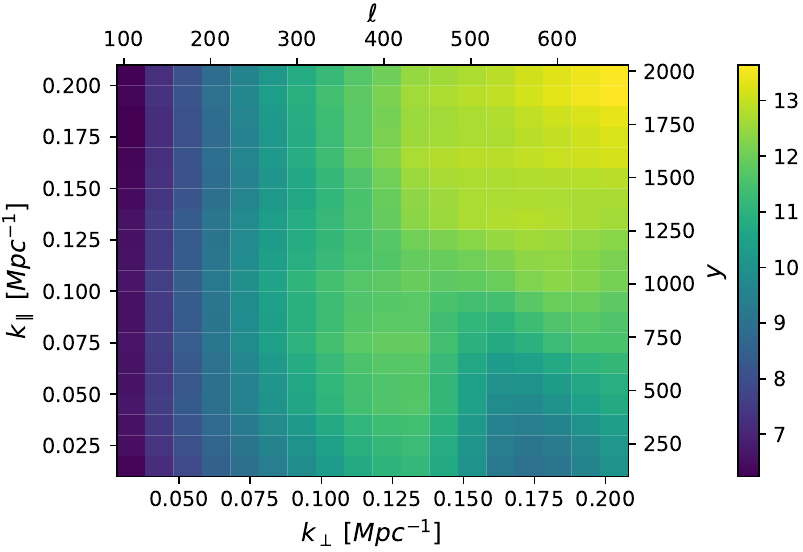}
\includegraphics[width=0.49\linewidth]{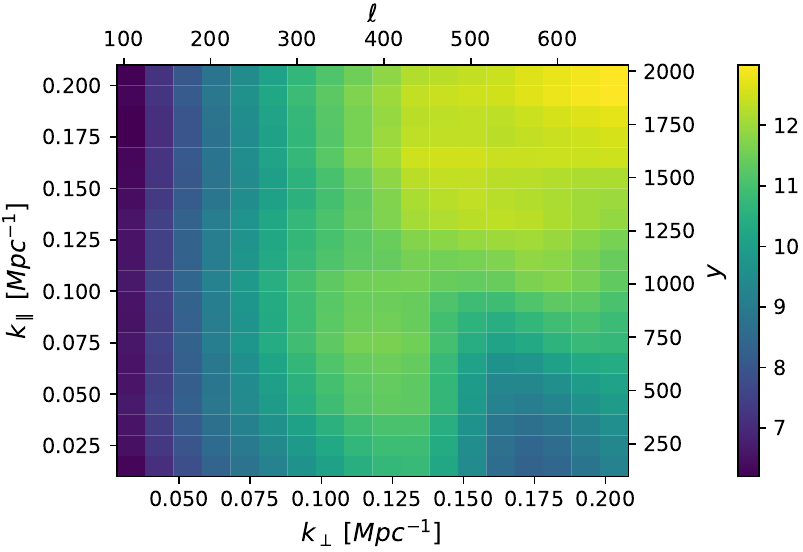}
\caption{In the upper panels we show the HI-HI-galaxy (\textit{Left})  and the HI-HI-shear (\textit{Right}) bispectrum signal in a patch against the gaussian contribution in the patch computed at the $z_i=0.95$ redshift bin. The gaussian contribution shows minimal variation with $y$ on large scales arising from the HI power spectrum variation in $y$. In the lower panels we show the the 2D signal-to-noise ratio for both cases computed in k-bins of 0.01 Mpc$^{-1}$ across the $k_\perp - k_\parallel$ plane for the redshift bin centered at $z_i=0.95$.}
\label{bispectrum}
\end{figure}

To quantify the statistical uncertainty of the cross-bispectrum, we compute its variance, $\mathcal{V}$, which over the full sky, is given as follows:
\begin{equation}
\begin{aligned}
\mathcal{V}\left[B_{\ell}^{\mathrm{sky}}(y;z_i)\right] &= (2\pi)^3 \delta_D^2(\boldsymbol{\ell} - \boldsymbol{\ell}') \delta_D(y-y')\langle B_{\ell}^{\mathrm{sky}} (y; z_i) B_{\ell}^{\mathrm{sky}} (y'; z_i) \rangle - \left( \langle B_{\ell}^{\mathrm{sky}} (y; z_i)  \rangle \right)^2  \\
&=  
2 \, \mathcal{V}[\dtone21 (\ell,y; z_i)]^2  
\mathcal{V}[\mathrm{X}( z_i)]  + {23}\, \left(B_{\ell}^{\mathrm{sky}}(y; z_i) \right)^2.
\end{aligned}
\label{bispec_var}
\end{equation}
The terms, $2 \, \mathcal{V}[\dtone21 (\ell,y; z_i)]^2  \mathcal{V}[\mathrm{X}(z_i)]$ and ${23}\, \left(B_{\ell}^{\mathrm{sky}}(y,z_i)\right)^2$, are the Gaussian and bispectrum squared (`BB') contributions, respectively. We neglect the term involving the cross-power spectrum, $\mathcal{V}[\dtone21 (\ell,y; z_i)] \left( C_\ell^{21,\mathrm{X}}(y; z_i) \right)^2,$ which is significantly reduced in amplitude after foreground removal. We also neglect the power spectrum-trispectrum (`PT') term, which is found to be negligible compared to the Gaussian contribution. In Figure \ref{shear_var_terms}, we show the variance terms for the HI-HI-galaxy density cross-bispectrum (solid curve) and HI-HI-cosmic shear cross-bispectrum (dashed curve) evaluated at a typical $y$. It is evident that the Gaussian contribution dominates the variance. In Figure \ref{bispectrum} ({\it{top panels}}), we compare the HI-HI-galaxy and HI-HI-shear cross-bispectrum signals in a given redshift bin to the Gaussian contribution, rescaled to the patch size. We note that the Gaussian noise contribution is subdominant to the cross-bispectrum signal for the scales we consider.

To assess the detectability of the cross-bispectrum, we compute the signal-to-noise ratio given in Equation (\ref{SNRbin}), with the signal $\mathcal{S}_i(\ell,y) = B_{\ell}^{\mathrm{sky}} (y; z_i)$ and the variance given in Equation (\ref{bispec_var}). In Figure \ref{bispectrum} ({\it{bottom panels}}), we see that the signal-to-noise ratio in the $k_\perp-k_\parallel$ plane indicates a strong detectability for the cross-bispectra measured by HIRAX and Rubin-LSST over a broad range of radial and transverse modes. Across the four redshift bins, we find that the cumulative signal-to-noise ratios are SNR=$[116, 208, 166, 29]$ for the HI-HI-galaxy density bispectrum and SNR=$[68, 202, 173, 36]$ for the HI-HI-cosmic shear bispectrum. For the HI-HI-galaxy density bispectrum, the signal-to-noise is higher at lower redshifts, where the galaxy density is higher. In contrast, the HI-HI-cosmic shear bispectrum has relatively more signal-to-noise at higher redshifts due to the broad redshift kernel of the cosmic shear. Both cross-bispectra have reduced signal-to-noise in the highest redshift bin due to relatively higher HI noise. It is worth noting that the cross-bispectrum signal-to-noise ratio is somewhat robust to foreground removal; we find that applying a pessimistic horizon-wedge foreground cut only degrades the signal-to-noise ratio by about 10\%. Having established the high signal-to-noise ratios of the auto-power spectra and cross-bispectra, we now investigate the parameter constraints that result from combining these spectra.

\section{Parameter Constraints}
\label{sec:Forecasts}
We forecast the astrophysical and cosmological parameter constraints achieved by combining the auto-power spectra and cross-bispectra probes. We use the Fisher formalism, where the Fisher matrix for a single probe, $Y \in  \left\{ C_\ell^{21}(y), C_\ell^\mathrm{X}(y) , B_\ell^{2121\mathrm{X}}(y) \right \},$ is given by \cite{tegmark1997measuring}
\begin{equation}
F_{ab}^{Y} = \frac{1}{2} S_{area} \Delta \tilde{\nu} \int \frac{d^2\ell}{(2\pi)^2} \int {dy \over (2\pi)} \partial_a (\ln Y ^{tot}) ~\partial_b (\ln Y^{tot}),
\label{eq:Fisher_matrix}
\end{equation}
with $Y^{tot}= Y^{signal} + Y^{noise}$ and $\partial_a$ the derivative with respect to the $a^{th}$ parameter. A more general treatment of the Fisher matrix would account for cross-covariance terms between the individual probes, with the Fisher matrix for each ($\ell,y$) mode given by:
\begin{equation}
F_{ij}(\ell,y) =  \partial_i \mathcal{Y} ^T \mathrm{Cov}^{-1} \partial_j \mathcal{Y}
\label{eq:Fisher_cov_per_ell_y}
\end{equation}
where
\begin{equation}
\mathcal{Y} = \left(     \begin{array}{c}
C_\ell^{21}(y)\\
B_\ell^{2121 \mathrm{X}}(y)\\
C_\ell^{ \mathrm{X}}(y)\\  
\end{array}       \right)
\end{equation} 
and $\mathrm{Cov}$ is the full covariance matrix.

We now demonstrate that the off-diagonal terms of the covariance matrix are negligible, such that we can sum the Fisher matrices of the individual probes using Equation \ref{eq:Fisher_matrix}. For compactness, we adopt the notation $P = C_\ell^{21}(y) ,$ $B = B_\ell^{2121\mathrm{X}}(y)$ and $X = C_\ell^{X}(y)$, to write the covariance matrix as
\begin{equation}
\mathrm{Cov} = \left(     \begin{array}{ccc}
\hat{C}_{PP} & \hat{C}_{PB} & \hat{C}_{X P}\\  
\hat{C}_{PB} & \hat{C}_{BB} & \hat{C}_{BX}\\
\hat{C}_{X P} & \hat{C}_{BX} & \hat{C}_{X X}
\end{array}       \right)
\end{equation}
The diagonal terms describe the variance of each field, that is, $\hat{C}_{PP} = ( C_{\ell}^{21} + C_{\ell}^{N,21} )^2$, $\hat{C}_{BB} = \mathcal{V}[B_{\ell}^{2121\mathrm{X}}]$ and $\hat{C}_{XX } = ( C_{\ell}^{X} + C_{\ell}^{N,X} )^2$. The HI cross-spectrum with galaxy density or cosmic shear, $\hat{C}_{XP} = 2 \mathcal{V}[C_\ell^{21,X}],$ becomes negligible after foreground removal. We compute the remaining terms using $\hat{C}_{PB} = \langle PB \rangle - \langle P\rangle \langle B \rangle$ and $\hat{C}_{BX} =  \langle BX \rangle - \langle B\rangle \langle X \rangle$. 
\begin{figure}[!t]
\centering
\includegraphics[width=0.6\linewidth]{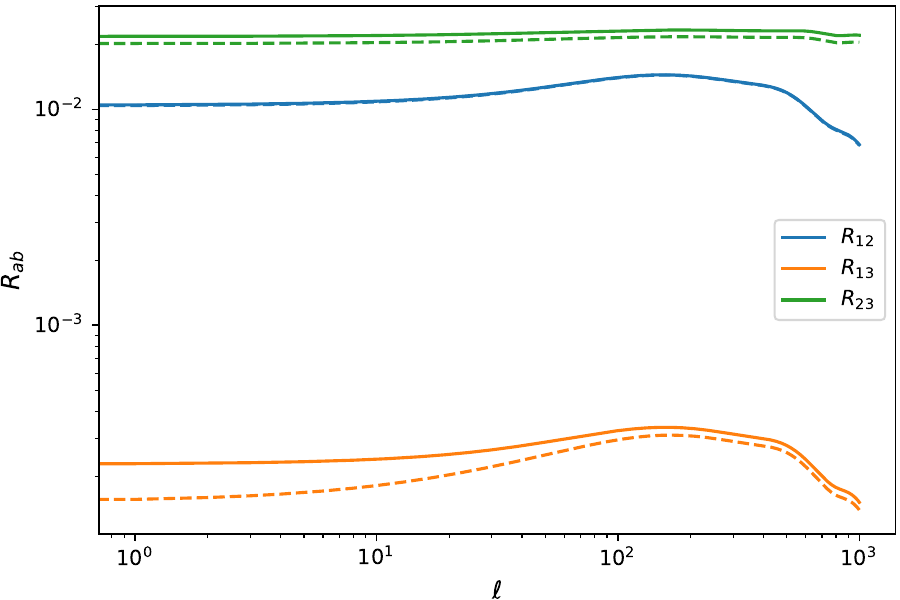}
\caption{Cross-covariance contributions, quantified by $R_{ab}$, to the Fisher matrix for the HI-galaxy density (solid curves) and HI-cosmic shear (dashed curves). The blue dashed and solid curves closely overlap.}
\label{Shear_cross_covariance_terms}
\end{figure}

To evaluate the contribution of the off-diagonal terms to the Fisher matrix, we consider the expansion of the Fisher matrix in Equation \ref{eq:Fisher_cov_per_ell_y}, which gives
\begin{equation}
\begin{aligned}
    F_{ab}(\ell,y) = {1\over \mathrm{det}[\mathrm{Cov}]}&\left[  \partial_{i}  P \partial_{j} P D_{11} + \partial_{i} B \partial_{j}  B D_{22} +\partial_{i}  X \partial_{j} X D_{33}  \right. \\ & \left. - 2 \partial_{i} P \partial_{j}  B D_{12} - 2\partial_{i}  P \partial_{j}  X D_{13} - 2 \partial_{i} B \partial_{j}  X D_{23}\right ],
\label{eq:Fisher_covariance_expanded}
\end{aligned}
\end{equation}
where $D_{11} = \hat{C}_{BB} \hat{C}_{XX } - \hat{C}_{BX}^2$, $D_{12}  = \hat{C}_{PB} \hat{C}_{XX } - \hat{C}_{BX}\hat{C}_{XP}$, $D_{13}  = \hat{C}_{PB} \hat{C}_{BX } - \hat{C}_{BB}\hat{C}_{XP}$, $D_{22} = \hat{C}_{PP} \hat{C}_{XX } - \hat{C}_{XP}^2$, $D_{23}  = \hat{C}_{PP} \hat{C}_{BX } - \hat{C}_{XP}\hat{C}_{PB}$ and $D_{33} = \hat{C}_{PP} \hat{C}_{BB } - \hat{C}_{PB}^2$. We define the dimensionless ratio, $R_{ab} = D_{ab}/\sqrt{D_{aa}D_{bb}},$ which will be small in the case of a negligible off-diagonal contribution.  
As seen in Figure \ref{Shear_cross_covariance_terms}, $R_{ab}$ is at most 2\% which allows us to neglect the off-diagonal contributions. Similar findings were reported for the cross-covariance between the halo power spectrum and squeezed limit halo bispectrum in \cite{biagetti2022covariance}.

\subsection{Redshift-Dependent and Model Parameters}
\label{sec:Redshift-Dependent and Model Parameters}
We first forecast the constraints on the redshift-dependent functions $\{ \sigma_8,\, b_\mathrm{HI}^{(1)},\, b_\mathrm{HI}^{(2)},\,  b_\mathrm{gal}, \, f\}$ and model parameters $\{ r_{\mathrm{HI\, HI\,}\deltag},\,  A_\mathrm{bao} \}$. For the redshift-dependent functions, we consider an independent value in each redshift bin.  We choose to fix $\Omega_{\mathrm{HI}},$ which is degenerate with $\sigma_8, \, b_{\mathrm{HI}}^{(1)}, b_{\mathrm{HI}}^{(2)}$ and $f$, but could, in principle, be constrained by a strong prior from external surveys \cite{holwerda2011looking, ponomareva2023mightee}. We include constraints on the amplitude of the BAO through the parameter, $A_\mathrm{bao},$ following the parameterization given in \cite{blake2003probing, Bull}.
The amplitude of the matter power spectrum, $\sigma_8$, is defined through the normalization
\begin{equation}
P_m(k) = \left({\sigma_8 \over \sigma_8^{fid}} \right)^2 P_m^{fid}(k).
\end{equation}
where the redshift-dependent amplitude is defined as $\sigma_8(z) = \sigma_8(z=0) D(z)$,

\begin{figure}[!t]
	\centering
	\includegraphics[width=1.0\linewidth]{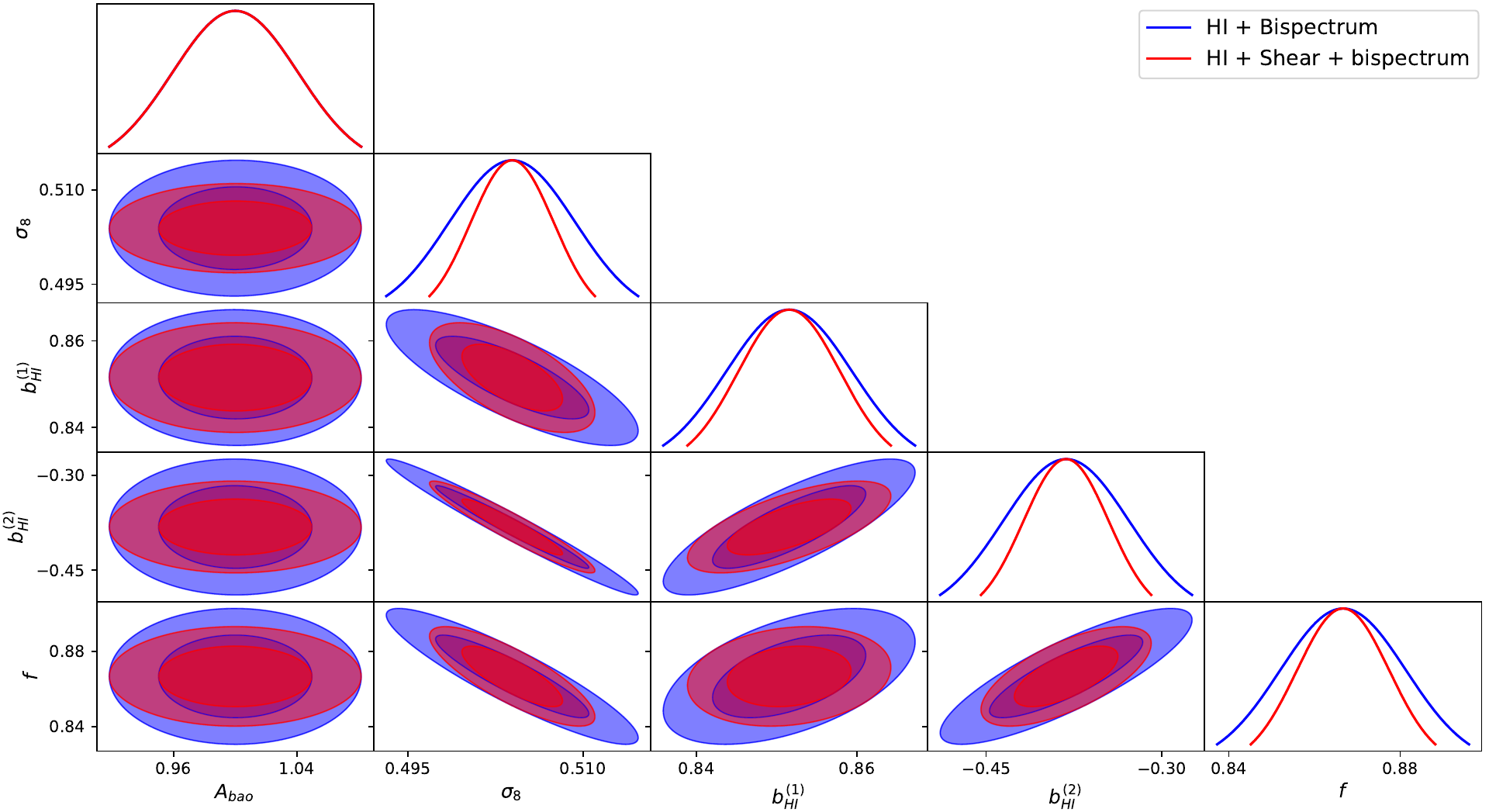}
	\caption{Forecast constraints on the redshift dependent and model parameters computed in the $z_i$=0.95 redshift bin from the HI-cosmic shear combination.}
	\label{fig:Bispec_shear_redshift_params}
\end{figure}
Many of the redshift-dependent and model parameters are degenerate within individual probes, but these degeneracies can be broken by combining probes. For example, the HI field depends on the combination of $\sigma_8 b_{\mathrm{HI}}^{(1)}$ and $\sigma_8 f,$ making $\sigma_8$ degenerate with $b_{\mathrm{HI}}^{(1)}$ and $f$.
The cosmic shear cross-bispectrum measures the combination of $\sigma_8$ with $b_{\mathrm{HI}}^{(1)}$, $b_{\mathrm{HI}}^{(2)}$ and $f$, however, it scales as $\sigma_8^4$ in the cosmic shear cross-bispectrum versus $\sigma_8^2$ in the HI power spectrum. This breaks the degeneracy of $\sigma_8$ with $b_{\mathrm{HI}}^{(1)}, \, b_{\mathrm{HI}}^{(2)} $ and $f$ when combining the HI power spectrum cosmic shear and cross-bispectrum.
\begin{figure}[!t]
	\centering
	\includegraphics[width=1.0\linewidth]{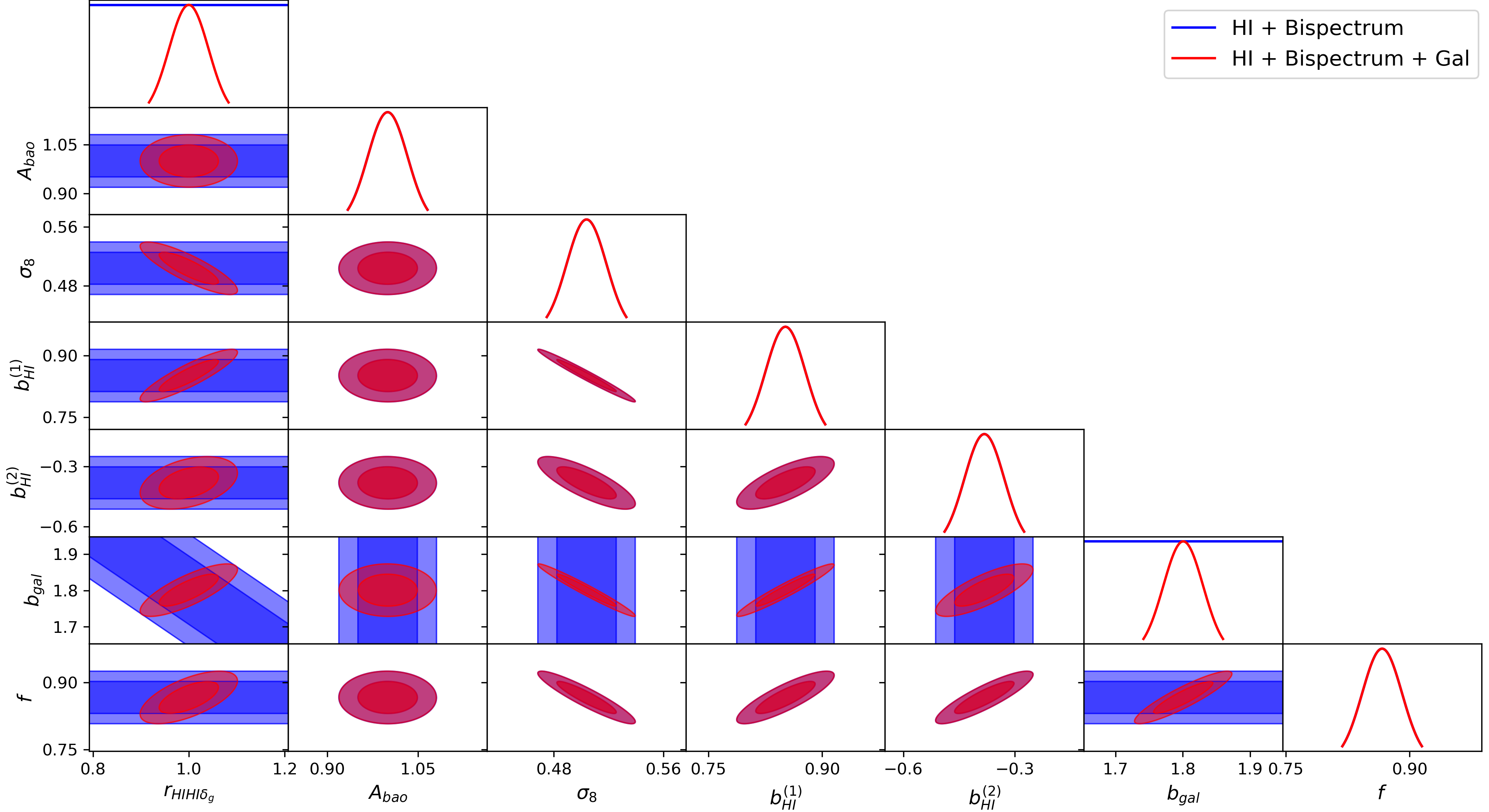}
	\caption{Forecast constraints on the redshift dependent and model parameters computed in the $z_i$=0.95 redshift bin from the HI-galaxy density combination.}
	\label{fig:Bispec_gal_redshift_params}
\end{figure}
In Figure \ref{fig:Bispec_shear_redshift_params}, we show the constraints that result from combining these probes in a single redshift bin, and in Table \ref{tab:redshift_constr} we list marginalized constraints across the four redshift bins. Adding the cosmic shear power spectrum further tightens the constraint on $\sigma_8,$ which in turn improves the constraints on $b_{\mathrm{HI}}^{(1)}$, $b_{\mathrm{HI}}^{(2)}$ and $f$. 

For the HI-galaxy density combination, the scenario is further complicated by the cross-correlation coefficient, $r_{\mathrm{HI\,HI\,}\deltag},$ and galaxy bias, $b_{\mathrm{gal}}$.  The combination of the HI power spectrum and galaxy density cross-bispectrum breaks the $\sigma_8$ degeneracy, allowing the galaxy density power spectrum to constrain $b_{\rm{gal}}$. This, in turn, breaks the degeneracy with the cross-correlation coefficient $r_{\mathrm{HI\, HI\, }\deltag}$, allowing us to constraints all parameters. In Figure \ref{fig:Bispec_gal_redshift_params}, we present the resulting marginalized parameter constraints from these probes in a single redshift bin, and in Table \ref{tab:redshift_constr} we list marginalized constraints across the four redshift bins. 

Across the four redshift bins presented in Table \ref{tab:redshift_constr}, the tightest constraint on $A_{\mathrm{bao}}$ of $3.1\%$ is obtained in the $z_i=1.27$ bin.  This constraint is obtained for both combinations: (i) the HI power spectrum, cosmic shear power spectrum, and cosmic shear cross-bispectrum (HI-cosmic shear combination); and (ii) the HI power spectrum, galaxy density power spectrum, and galaxy density cross-bispectrum (HI-galaxy density combination). In the following subsections, we discuss, in turn, the constraints on the structure growth parameters, HI and galaxy biases, and the HI-HI-galaxy cross-correlation coefficient resulting from these probes.

\begin{table}[!ht]
	\centering
    \resizebox{\textwidth}{!}{
	\begin{tabular}{|c|c|c|c|c|c|c|c|c|}
		\hline
		\multirow{2}{*}{} & \multirow{2}{*}{Redshift bin center} & \multicolumn{7}{c|}{Parameters} \\ \cline{3-9}
		&  & $r_{\mathrm{HIHI}\deltag}$ & A$_\mathrm{bao}$ & $\sigma_8$ & $b_\mathrm{HI}^{(1)}(z)$ & $b_\mathrm{HI}^{(2)}(z)$ & $b_\mathrm{gal}(z)$&  $f(z)$ \\
		\hline
		
		\multirow{4}{*}{HI-cosmic shear combination}
		& $0.81$ & -- & 0.077 & 0.011 &  0.015 &  0.10 &-- &0.028\\ \cline{2-9}
		& $0.95$ & -- & 0.033 & 0.0029 & 0.0052 &  0.030 &-- &  0.011 \\ \cline{2-9}
		& $1.27$ & -- &  0.031 &  0.0014 & 0.0042  &  0.020&-- & 0.0074\\ \cline{2-9}
		& $1.95$ & -- & 0.087 & 0.00073 & 0.0082 & 0.040 &-- &0.012 \\

        \hline
		
		\multirow{4}{*}{HI-galaxy density combination}
		& $0.81$ & 0.080  & 0.063 & 0.035 & 0.056 & 0.12 & 0.069 &  0.054\\ \cline{2-9}
		& $0.95$ & 0.041  & 0.033  & 0.015 & 0.026 & 0.054 & 0.030& 0.024 \\ \cline{2-9}
		& $1.27$ & 0.037 & 0.031 & 0.011 &  0.025 &0.057 &   0.026&0.023 \\ \cline{2-9}
		& $1.95$ & 0.13 & 0.095 &0.035  & 0.11 & 0.29 &0.102 &  0.098\\
	
		\hline
	\end{tabular}
    }
	\caption{Constraints on the redshift-dependent and model parameters from: (i) the HI power spectrum, cosmic shear power spectrum, and cosmic shear cross-bispectrum (HI-cosmic shear combination); and (ii) the HI power spectrum, galaxy density power spectrum, and galaxy density cross-bispectrum (HI-galaxy density combination). }
    \label{tab:redshift_constr}
\end{table}

\subsubsection{Structure Growth Parameters}
Here we discuss the constraints on the growth parameters, $\sigma_8$ and $f,$ presented in Table \ref{tab:redshift_constr}. The best constraints on $\sigma_8$ and $f$ of $0.073\%$ and $0.74\%,$ respectively, are from the HI-cosmic shear combination. The cosmic shear constraints on $\sigma_8$ improve with higher redshift and with larger redshift bins. This leads to better constraints on $b_{\mathrm{HI}}^{(1)}$ and $f$ in the higher redshift bins, up to the highest redshift bin where the HI constraining power decreases. For the HI-galaxy density combination the best constraints on $\sigma_8$ and $f$ are $1.1\%$ and $2.3\%,$ respectively. The HI-galaxy density constraining power peaks in the central bins and weakens in the highest redshift bin, due to the declining galaxy number density at higher redshifts.

\subsubsection{Bias Parameters}
The combination of probes provides precise constraints on the bias parameters. From Table \ref{tab:redshift_constr}, the tightest constraints on the HI biases, $b_{\mathrm{HI}}^{(1)}$ and $b_{\mathrm{HI}}^{(2)}$, are $0.42\%$ and $2.0\%$, respectively, from the HI-cosmic shear combination. From the HI-galaxy density combination, the tightest constraint on the galaxy density bias, $b_{\mathrm{gal}},$ is $2.6\%$. Such precise measurements probe the clustering of neutral hydrogen and galaxies relative to dark matter, and allow us to retain competitive cosmological constraints when marginalizing over these biases. Moreover, the constraints on $b_{\mathrm{HI}}^{(2)},$ enabled by the cross-bispectrum measurement, allow us to constrain the nonlinear contributions to the clustering signal. Simultaneous constraints on the linear and nonlinear HI biases are necessary because nonlinear corrections modify their effective amplitudes \cite{sarkar2019modelling,penin}. 

\begin{figure}[!t]
    \centering
    \begin{subfigure}
        {\includegraphics[width=0.48\linewidth]{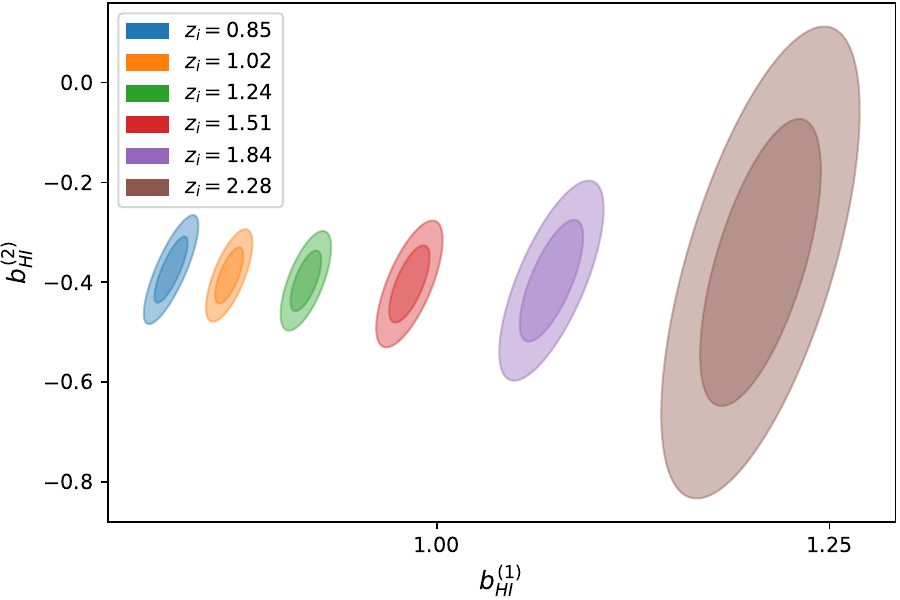}}
    \end{subfigure}
    \begin{subfigure}
        {\includegraphics[width=0.48\linewidth]{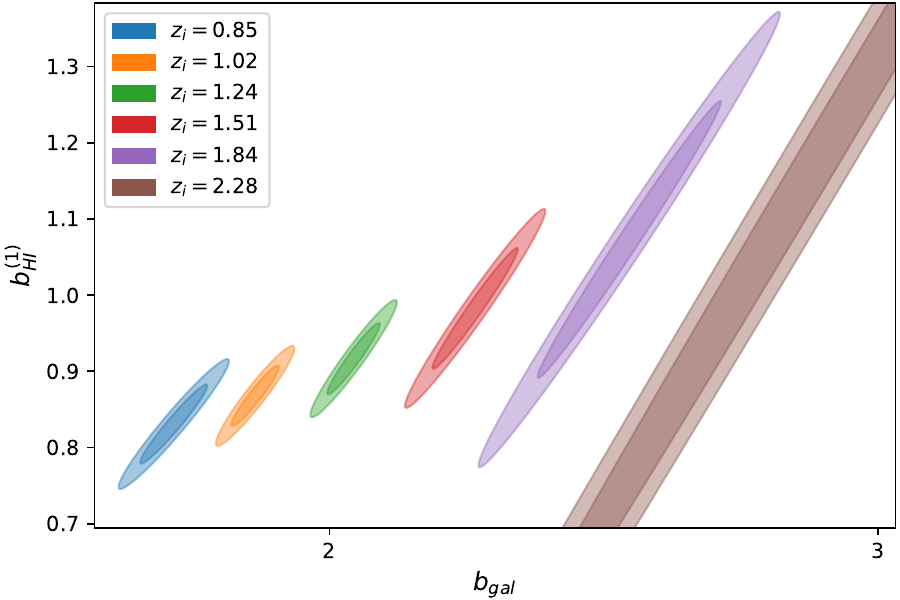}}
    \end{subfigure}
    \caption{Forecast constraints on the HI bias parameters in six redshift bins of width, $\Delta\nu \sim 66.67\mathrm{MHz} ,$ from the HI-cosmic shear combination (\textit{left}). Forecast constraints on the first-order HI bias and the galaxy density bias from the HI-galaxy density combination for the same six redshift bins (\textit{right}).}
    \label{fig:Bias_evolution_constraints}
\end{figure}

To further demonstrate the constraining power of these probes on the biases as a function of redshift, we consider the constraints in six redshift bins of width $66.67$ MHz. We show in Figure \ref{fig:Bias_evolution_constraints} (\textit{left}), the constraints on the first- and second-order HI biases from the HI-cosmic shear combination, which provides the tightest constraints on these parameters. We obtain constraints of 0.6\% to 2.6\% on, $b_{\mathrm{HI}}^{(1)},$ and 3.8\% to 19\% on $b_{\mathrm{HI}}^{(2)},$ across all six redshift bins.
We observe that we obtain relatively uniform and precise constraints across the redshift bins, which will be useful for constraining models in which the bias evolves.

We next consider the constraints from the HI-galaxy density combination on the first-order HI and galaxy biases 
in the six redshift bins, as shown in Figure \ref{fig:Bias_evolution_constraints} (\textit{right}). The highest redshift bin yields a weak constraint on both parameters (59\% on $b_{\mathrm{HI}}^{(1)}$ and 49\% on $b_{\mathrm{gal}}$) since the HI noise and galaxy shot noise increase at high redshift. The constraints on $b_{\mathrm{HI}}^{(1)}$ range from, 2.7\% to 12\% across the first five redshift bins, while the constraints on $b_{\mathrm{gal}}$ range from 2.9\% to 11\%.

\subsubsection{Cross-Correlation Parameter}
The cross-correlation coefficient probes the stochasticity between HI gas and the distribution of galaxies. In the absence of stochasticity and scale-dependent bias, the probes trace the same matter density field, resulting in $r_{\mathrm{HI\,HI\,}\deltag}=1$ on sufficiently large scales. Environmental and astrophysical processes lead to deviations in $r_{\mathrm{HI\,HI\,}\deltag}$ from unity, therefore, constraints on the cross-correlation coefficient probe these processes. The tightest constraint, over the four redshift bins, we achieve for the constant cross-correlation coefficient is 3.7\% in the $z_i=1.27$ bin as seen in Table \ref{tab:redshift_constr}. Over the four redshift bins the constraints range from 3.7\% to 13\% providing an excellent opportunity to probe its scale dependence.

While the constant cross-correlation coefficient is expected to be approximately unity on linear scales, numerical simulations and semi-analytic models indicate a scale dependence at quasi-linear and nonlinear scales \cite{villaescusa2018ingredients, zhou2024parametrization}. This scale dependence is well described by various parameterizations, with a quadratic model providing an accurate description over relevant cosmological scales \cite{zhou2024parametrization}. Motivated by this and the precise constraints obtained on the constant cross-correlation coefficient, we consider a phenomenological model of the following form:
\begin{equation}
    r_{\mathrm{HI\,HI\,}\deltag}(k) = r_0 + r_1 k.
\end{equation}

The first-order model is sufficient since we only consider linear scales and do not probe the quadratic order corrections. Furthermore, our aim is to assess scale-dependent deviations from $ r_{\mathrm{HI\,HI\,}\deltag} = 1,$ not to reconstruct the exact functional form. In the above parameterization, $r_0$ represents the constant large-scale correlation and $r_1$ parametrizes the linear scale dependence of the cross-correlation coefficient. The fiducial model adopts $r_0 = 1$ with $r_1$ chosen such that the cross-correlation coefficient decreases by 5\% on mildly nonlinear scales, i.e., $ r_{\mathrm{HIHI}\deltag}(k = 0.14\, \mathrm{Mpc}^{-1}) = 0.95$. In principle, $r_1,$ could be redshift dependent if one normalizes it at the redshift-dependent nonlinear scale, $k_{NL}(z)$. Since our objective is to demonstrate the detectability of generic scale-dependent deviations, we adopt a single redshift-independent value of $r_1 = -0.357$ across all redshift bins. 

\begin{figure}[!t]
    \centering
    \includegraphics[width=0.7\linewidth]{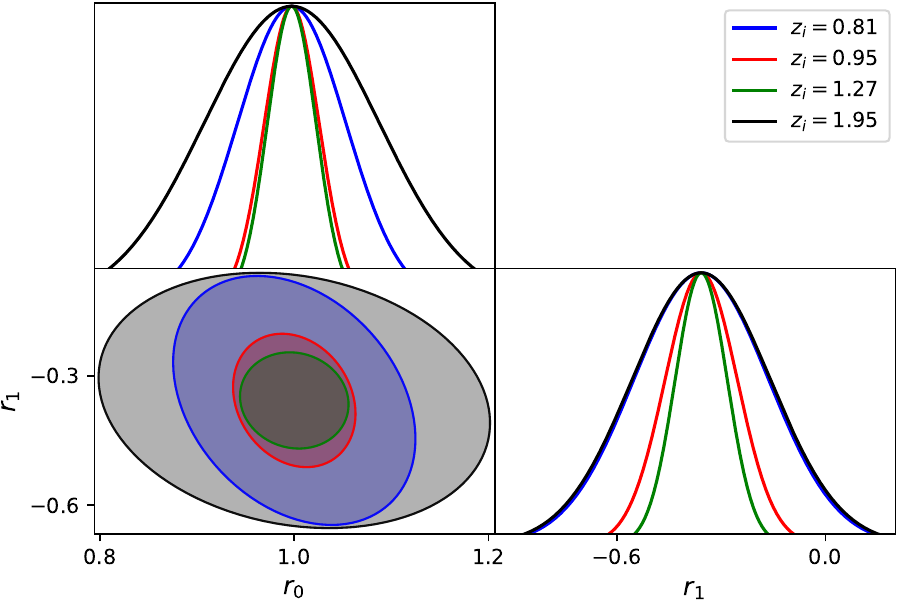}
    \caption{The 1-$\sigma$ forecast constraints on the scale-dependent parametric form of the cross-correlation coefficient computed at the four logarithmically chosen redshift bins.}
    \label{fig:Cross_correlation_coeff}
\end{figure}

The constraints on $r_0$ and $r_1$ are shown in Figure \ref{fig:Cross_correlation_coeff}. The constraints over the four redshift bins range from $3.8\%$ to $13.5\%$ on $r_0$ and 7.5\% to 19.9\% on $r_1$. The constraints improve with increasing redshift up to the highest redshift bin, where the HI noise and galaxy shot noise increase. The increased range of accessible modes at higher redshifts reduces the degeneracy between $r_0$ and $r_1$. The constraints on $r_1$ provide a meaningful probe of the scale dependence in the stochasticity between HI gas and galaxies on quasi-linear scales.

\subsection{Cosmological Parameters}

We now investigate constraints on the cosmological model through the Alcock-Paczynski parameters \cite{alcock1979evolution}. We first extend the Fisher matrix in Section \ref{sec:Redshift-Dependent and Model Parameters} to include the distance scale parameters \cite{Bull,blake2003probing}
\begin{equation}
\alpha_\perp = { \cpl^{fid} \over \cpl } = {D_A^{fid}(z) \over D_A(z)}, \hspace{2mm}
\alpha_\parallel = { r_\nu^{fid} \over r_\nu} = {H(z) \over H^{fid}(z)},
\end{equation}
by substituting $\ell \rightarrow \alpha_\perp \ell$ and $y \rightarrow \alpha_\parallel y$. Here the angular diameter distance, $D_A(z),$ and the Hubble expansion rate, $H(z),$ probe the transverse and radial distances, respectively. The parameter set is then updated by marginalizing over the biases, BAO amplitude and cross-correlation coefficient in each redshift bin. Here we constrain the amplitude of the power spectrum at present time, $\sigma_8(z=0)$. We denote the resulting redshift-dependent Fisher matrix as, $F_{a b}^{(i)},$ in each redshift bin, $i$, with parameters $p = \{ \sigma_8, f, \alpha_\perp, \alpha_\parallel \}.$

We then construct the redshift-dependent Fisher matrix, $F'^{(i)}_{a' b'}$, that contains cosmological parameters, $p' = \{ \Omega_{m}, \sigma_8, h, n_s, \omega_b, w_0, w_a \},$ by transforming $F_{a b}^{(i)}$ above as follows:
\begin{equation}
\left[ F'^{(i)}_{a' b'} \right] = \left[ M^{(i)}_{a a'} \right]^T \left[ F^{(i)}_{a b} \right] \left[ M^{(i)}_{b b'} \right].
\end{equation}
Here, the matrix, $M^{(i)}_{a a'}={\partial_{p_a} / \partial_{p'_{a'}}},$ transforms $F^{(i)}_{a b}$ to $F'^{(i)}_{a' b'}.$ We have included the spectral index of primordial density perturbations, $n_s,$ and the baryon density, $\omega_b,$ in the cosmological parameters. The baryon density is not directly constrained by HI surveys, but is strongly correlated with other parameters constrained by the Planck prior. The density parameters satisfy the dimensionless Friedmann equation
$
\Omega_{\mathrm{DE}} + \Omega_{m} = 1,
$
where we assume a spatially flat universe, $\Omega_k = 0$. 

The overall constraints on the cosmological parameters are obtained from the Fisher matrix resulting from the sum of the Fisher matrices,  $F'^{(i)}_{a' b'}$ , over the four redshift bins. The HI power spectrum and the cross-bispectrum are found to be independent between bins, resulting in a diagonal bin-bin covariance. For the cosmic shear power spectrum, we include the cross-covariances of tomographic bins arising from the overlap of the lensing kernels $W_{\gamma}(\chi_i)$ and $W_{\gamma}(\chi_j)$. \footnote{This is because galaxies in different source bins are lensed by the same foreground density fluctuations, and hence, the cross-power spectra $C_\ell^{\gamma,ij}$ are non-zero and contribute to the covariance.} In the following subsections, we compute constraints for the $\Lambda$CDM and $w_0w_a$CDM models. We adopt Planck 2018 priors \cite{Aghanim} on the cosmological parameters, with $\tau$ pre-marginalized.

\subsubsection{$\Lambda$CDM Parameters}
\begin{figure}[!t]
    \centering
    \includegraphics[width=1.0\linewidth]{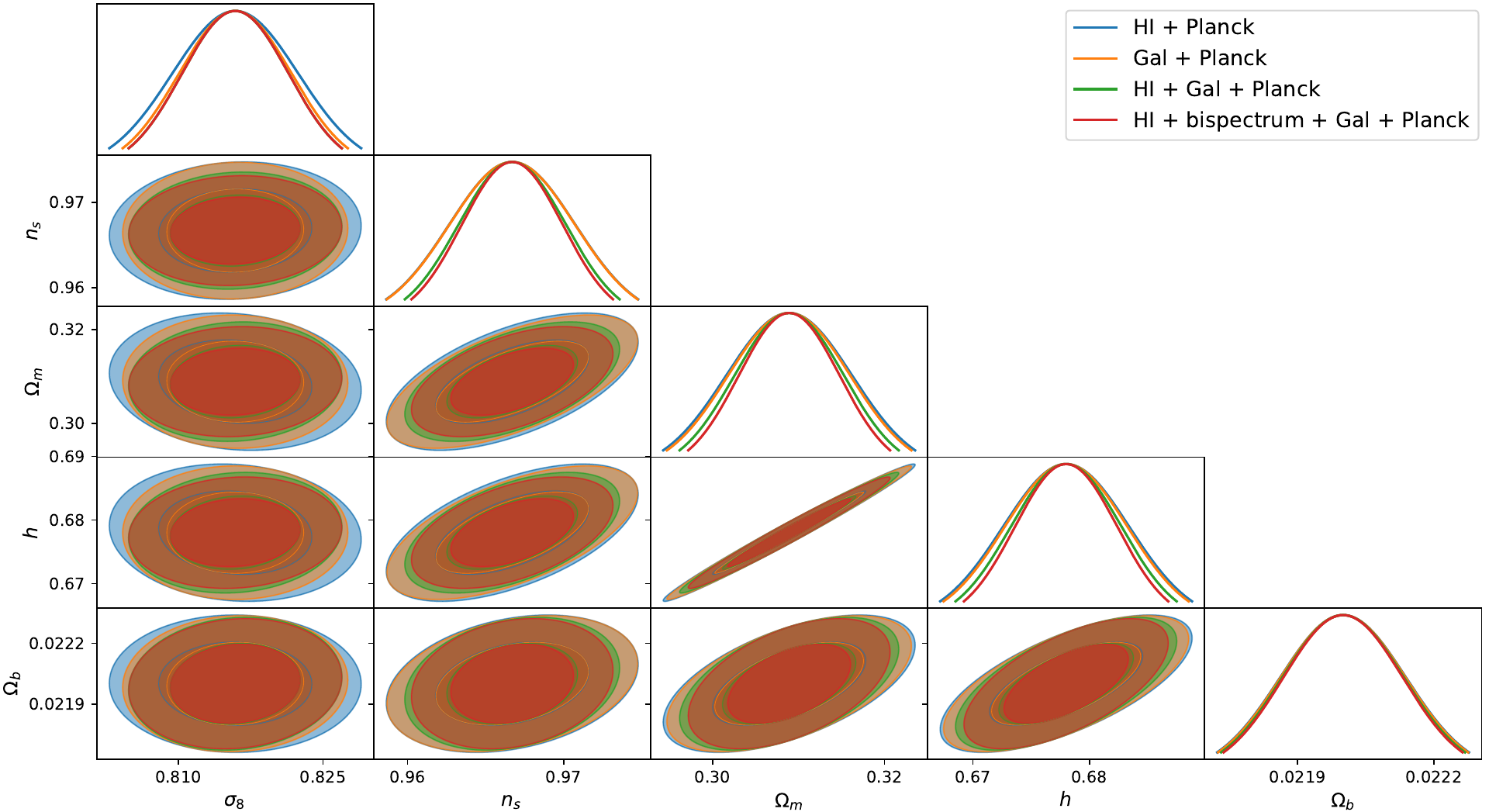}
    \caption{Forecast constraints on the $\Lambda$CDM model from the HI-galaxy density combination.}
    \label{fig:Gal_lcdm_constraints}
\end{figure}

Here we present constraints on the $\Lambda$CDM parameters, $\{\Omega_m, \sigma_8, h, n_s, \omega_b\},$ by fixing $w_0$ and $w_a$. 
\begin{figure}[!t]
    \centering
    \includegraphics[width=1.0\linewidth]{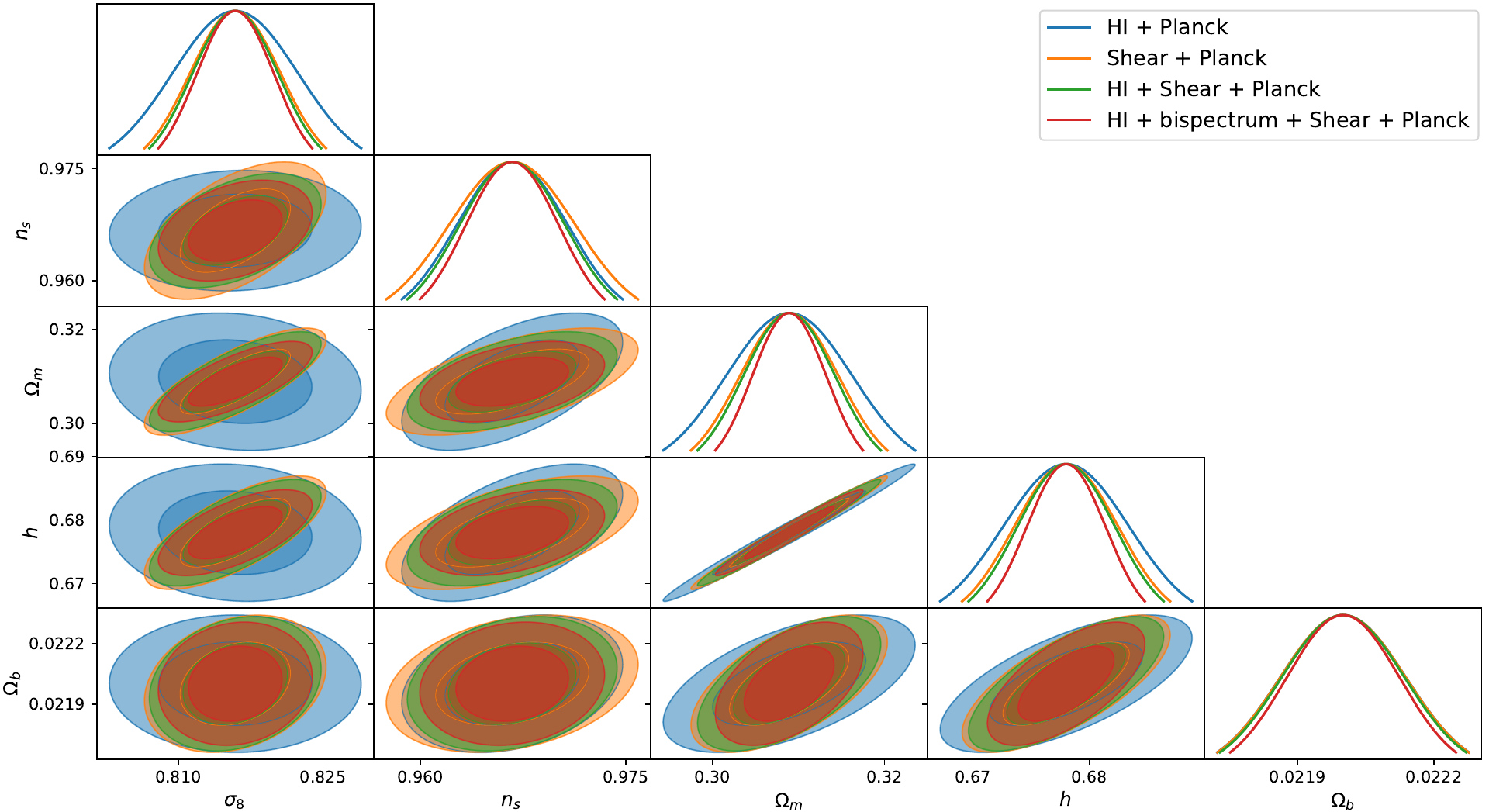}
    \caption{Forecast constraints on the $\Lambda$CDM model from the HI-cosmic shear combination.}
    \label{fig:Shear_lcdm_constraints}
\end{figure}
The resulting constraints are shown in Figures \ref{fig:Shear_lcdm_constraints} and \ref{fig:Gal_lcdm_constraints} for the HI-cosmic shear and HI-galaxy density combinations, respectively. We see that including the cross-bispectrum improves the constraints obtained from the combination of the HI and galaxy density or cosmic shear power spectra. This improvement is stronger for the HI-cosmic shear combination, where the cosmic shear cross-bispectrum has one less bias parameter and no cross-correlation coefficient parameter. Consequently, as shown in Table \ref{tab:LCDM_errors}, the HI-cosmic shear combination yields the strongest constraints on $h,$ $\sigma_8$ and $\Omega_m$ of $0.28\%$, $0.31\%$ and $0.35\%,$ respectively. The HI-galaxy density combination also achieves sub-percent constraints on the $\Lambda$CDM model, providing the tightest constraint of $0.26\%$ on $n_s$. The figure of merit for $\sigma_8$ and $\Omega_m,$ defined as $\mathrm{FoM}_{\Omega_m - \sigma_8} = [\det \mathrm{Cov}(\Omega_m, \sigma_8)]^{-1/2}$ \cite{abbott2018dark}, is $125632$ for the HI-cosmic shear combination and $84477$ for the HI-galaxy density combination. In comparison, 
the DES Y6 using 3$\times$2pt with low-redshift \footnote{DESI DR2 BAO, DES BAO (excl DESI), DES
SN, SPT CL.} and the CMB \footnote{ACT DR6, SPT-3G DR1 (primary probes, not including CMB lensing)}, the figure of merit is 51955 \cite{descollaboration2026darkenergysurveyyear}.

\begin{table}[!htb]
	\centering
    \resizebox{\textwidth}{!}{
	\begin{tabular}{|c|c|c|c|c|c|}
		\hline
		& $\Omega_m$ &$\sigma_8$& $h$ & $n_s$  &$\Omega_b$\\
		\hline
		\textit{Planck} & 0.0074 & 0.0060 & 0.0054 & 0.0042 & 0.00015\\
		\hline
		\text{HIRAX} + \textit{Planck}& 0.0060 & 0.0053 & 0.0044 &  0.0033 & 0.00014\\
		\hline
            \text{Gal} + \textit{Planck}& 0.0058& 0.0048  &0.0043 & 0.0033 & 0.00014   \\
		\hline
            \text{Shear} + \textit{Planck}& 0.0047 & 0.0038& 0.0036 & 0.0038  & 0.00014 \\
		\hline
          \text{HIRAX} +  \text{Shear} + \textit{Planck}& 0.0044 & 0.0036& 0.0034& 0.0031  & 0.00014 \\
		\hline
		\text{HIRAX} + \text{Shear-Bispectrum} + \textit{Planck} & 0.0037 &0.0043 & 0.0028 &0.0028   & 0.00013\\
		\hline
		\text{HIRAX} + \text{Shear-Bispectrum} +  \text{Shear} + \textit{Planck} & 0.0035& 0.0031 & 0.0028 & 0.0028 &0.00012 \\
		\hline
          \text{HIRAX} +  \text{Gal} + \textit{Planck}& 0.0052 & 0.0045& 0.0039& 0.0028  & 0.00013 \\
		\hline
		\text{HIRAX} + \text{Gal-Bispectrum} + \textit{Planck}  &  0.0054 &0.0053 &  0.0040 &0.0030 & 0.00013\\
		\hline
		\text{HIRAX} + \text{Gal-Bispectrum} + \text{Gal} + \textit{Planck}  & 0.0050 & 0.0045 &  0.0039 & 0.0026 & 0.00013\\
		\hline
	\end{tabular}
    }
	\caption{ Marginalized 68\% $\Lambda$CDM parameter forecast constraints for the HIRAX experiment alone and in combination with the galaxy density and the cosmic shear auto-correlation and cross-bispectra. }
	\label{tab:LCDM_errors}
\end{table}

\subsubsection{$w0wa$CDM Parameters}
To include constraints on the dark energy equation-of-state parameters, we adopt the CPL parameterization \cite{albrecht2006report}
\begin{equation}
w(z) = w_0 + w_a {z\over 1+z},
\end{equation}
and constrain the parameters $w_0$ and $w_a$. Including the equation-of-state parameters degrades the $\Lambda$CDM constraints. This marginalization primarily affects distance-scale parameters, while parameters such as $n_s$ and $\sigma_8$, are less affected \cite{Bull}. Therefore, following marginalization, we focus on $\Omega_{\mathrm{DE}}$ (obtained from $\Omega_{\mathrm{DE}} + \Omega_{m} = 1$)  and $h$ which exhibit the strongest correlations with $w_0$ and $w_a$. For these constraints, we apply a horizon wedge foreground cut, in addition to the $k_\parallel<k_{FG}$ cut, to the HI and cross-bispectrum signals and find that the constraints degrade by a few percent (typically 3-5\%) across parameters. These results demonstrate that the constraints remain largely robust to the removal of the horizon wedge modes, even with the inclusion of dark energy parameters.

\begin{figure}[!t]
	\centering
	\begin{subfigure}
		{\includegraphics[width=0.49\linewidth]{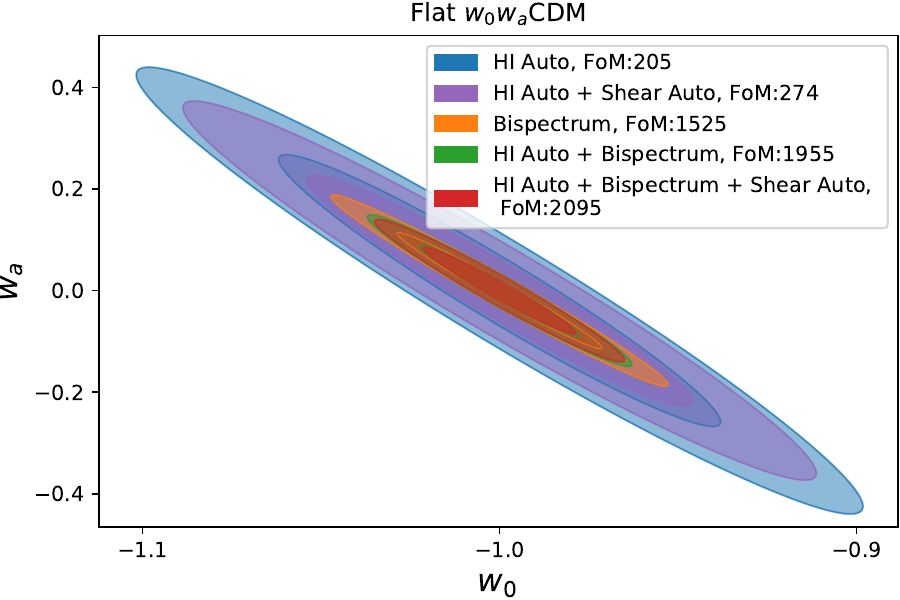}}
	\end{subfigure}
	\begin{subfigure}
		{\includegraphics[width=0.49\linewidth]{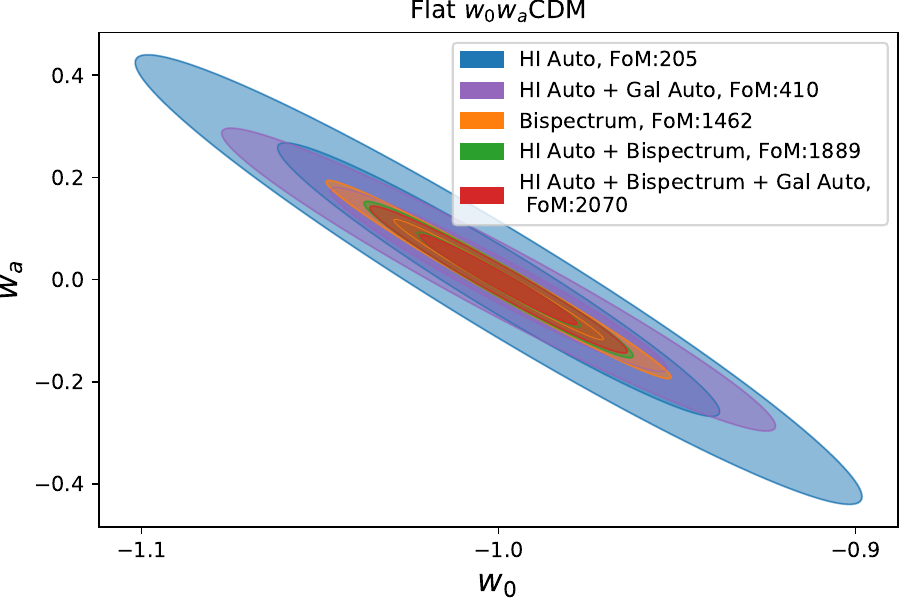}}
	\end{subfigure}
	\caption{Forecast constraints on dark energy equation of state parameters combined over all redshift bins. We show the HI-cosmic shear combination case (\textit{left}) and the HI-galaxy density combination (\textit{right}).}
	\label{fig:DE_constraints}
\end{figure}

The marginalized constraints that show the improvement from the combined probes on dark energy constraints are summarized in Table \ref{tab:Cosmo_DE_errors_full}. The best constraints on $\Omega_{\rm DE}$ and $h$, obtained from the HI-cosmic shear combination, are 0.21\% and 0.15\%, respectively. We obtain 1.4\% and 1.5\% constraints on $w_0$ and 5.7\% and 5.9\% on $w_a,$ from the HI-cosmic shear and HI-galaxy density combinations, respectively. This constraint is improved significantly by the addition of the cross-bispectrum to the power spectra combination, which are a factor of 2 to 3 worse than the full combination of probes. In Figure \ref{fig:DE_constraints} we show the constraints on the dark energy equation-of-state parameters for the HI-cosmic shear \textit{(left)} and HI-galaxy density (\textit{right}) combinations. We find that adding the cross-bispectrum improves the dark energy FoM of the HI only case from 205 to 1955 and 1889 for the HI-cosmic shear and HI-galaxy density combinations, respectively. The addition of the cross-bispectrum improves the power spectra combinations from 274 and 410 to 2095 and 2070 for the HI-cosmic shear and HI-galaxy density combinations, respectively.


\begin{table}[!htb]
	\centering
    \resizebox{\textwidth}{!}{
	\begin{tabular}{|c|c|c|c|c|}
		\hline
		 & $\Omega_{\rm DE}$ & $w_0$ & $w_a$ &$h$\\
		\hline
		\text{HIRAX} + \textit{Planck} & 0.0041 & 0.042 & 0.18 & 0.0027 \\
		\hline
		\text{HIRAX}  +  \text{Shear} + \textit{Planck}&0.0029 & 0.036 & 0.15& 0.0020  \\
		\hline
		\text{HIRAX} + \text{Shear-Bispectrum} +  \text{Shear} + \textit{Planck}&0.0021 & 0.014 & 0.057& 0.0015  \\
		\hline
		\text{HIRAX} +  \text{Gal} + \textit{Planck}& 0.0036& 0.032&  0.12&  0.0025  \\
		\hline
		\text{HIRAX} + \text{Gal-Bispectrum} +  \text{Gal} + \textit{Planck}& 0.0024& 0.015&  0.059&  0.0017  \\
		\hline
	\end{tabular}
    }
	\caption{\label{tab:Cosmo_DE_errors_full} Marginalized 68\% $w0wa$CDM parameter forecast constraints for the HIRAX experiment alone and in combination with the galaxy density and the cosmic shear auto-correlation and cross-bispectra.
	}
\end{table}

\section{Discussion}
\label{sec:conclusion}

In this work, motivated by the vanishing of the cross-correlation signal between HI and galaxy density or cosmic shear probes after foreground removal, we proposed and studied a cross-bispectrum. The cross-bispectrum is based on the correlation of two HI modes with a background galaxy density or cosmic shear mode. We considered HIRAX for the HI measurements and Rubin LSST surveys for the cosmic shear and galaxy density measurements. We found that the cross-bispectrum is measured with high signal-to-noise in a given redshift bin by these surveys. This signal-to-noise is shown to be somewhat robust to foreground removal.

We derived constraints on redshift-dependent and model parameters from combining these probes. The best constraints achieved on $f$ and $\sigma_8$ using the HI-cosmic shear combination are $0.74\%$ and $0.073\%$, respectively. Current constraints from galaxy clustering and galaxy-galaxy lensing combinations typically achieve errors of around CHECK 0.22 on $f$ and 0.06 on $\sigma_8$ \cite{de2017vimos, jullo2019testing}. Our constraints improve the $\sim 4\%$  constraints on these parameters reported in \cite{gil2016clustering}, where the BOSS galaxy power spectrum and bispectrum were combined to break the $f$-$\sigma_8$ degeneracy. Forecasts, using the galaxy power spectrum with the galaxy bispectrum or phase-based correlations, presented in \cite{byun2020constraining} and \cite{perenon2020improvements} report $\sim 3\%$ and $\sim 1\%$ errors on $f$, and $\sim 0.5\%$ and $\sim 1\%$ errors on $\sigma_8$, respectively. DESI clustering measurements of the full shape galaxy power spectrum have also provided a constraint of 3.4\% on $\sigma_8$ \cite{adame2025desi}. 

We also studied constraints on the HI and galaxy biases, across our redshift bins, obtaining the best constraints of 0.42\%, 2.0\% and 2.6\% on $b_{\mathrm{HI}}^{(1)},$ $b_{\mathrm{HI}}^{(2)},$ and $b_{\mathrm{gal}},$ respectively, in a given bin. We also constrained the HI-galaxy density cross-correlation coefficient, $r_{\mathrm{HI\,HI\,}\deltag}(k) = r_0 + r_1k,$ and its quasi-linear scale dependence, obtaining the tightest constraints of 3.8\% and 7.5\% on $r_0$ and $r_1,$ respectively, in a given redshift bin. Measurements of HI and galaxy biases, together with the cross-correlation coefficient over cosmological redshift ranges remain relatively limited. In particular, cross-correlation measurements primarily constrain combinations involving, $b_{\rm HI} r_{\rm HIg}$ \cite{masui2013measurement, wolz2022h, cunnington2023h}. Here we are able to break the degeneracies and provide constraints on the first and second order HI bias, the galaxy bias and the quasi-linear scale dependence of cross-bispectrum cross-correlation coefficient. 

We then studied constraints on the cosmological model. For the $\Lambda$CDM model, we obtained constraints of 0.31\% and 0.35\% on $\sigma_8$ and $\Omega_M$ respectively, using the HI-cosmic shear combination. This yields a figure of merit of 125632 which improves the result of the DES Y6 3$\times$2pt with external low-redshift and CMB data by a factor of 2.4. We obtained constraints of 2.6\%, 0.28\% and 0.12\% on $n_s$, $h$ and $\omega_b$ respectively. 
For the $w0wa$CDM model, we found that we can constrain the dark-energy equation-of-state parameters at the 1.4\% and 5.7\% level for $w_0$ and $w_a,$ respectively, from the HI-cosmic shear combination. Recent measurements from the eBOSS cosmology analysis, combined with Planck, Pantheon SNe Ia, and DES Y1, yield approximately 7\% and 80\% constraints on $w_0$ and $w_a$, respectively \cite{alam2021completed}, while an analysis of DESI BAO in combination with CMB data provides constraints, $w_0 = -0.45^{+0.34}_{-0.21}$ and $w_a = -1.79^{+0.48}_{-1.0}$ \cite{adame2025desi}.
Our results improve the dark energy FoM forecasts for SKA \cite{bacon2020cosmology} and DESI \cite{vargas2019unraveling} by a factor of 7 or more. Forecasts for a joint SKA1 analysis with Planck priors yields 7\% and 34\% constraints on $w_0$ and $w_a,$ respectively,  \cite{bacon2020cosmology}, while Euclid forecasts yield 2\% and 7\% \cite{ilic2022euclid}. 

Our analysis assumes idealized conditions, including perfect calibration without systematic errors. We also neglect shear systematics, such as redshift calibration uncertainties or intrinsic alignment effects, which degrade the constraining power of weak-lensing surveys. We have neglected nonlinear corrections to the bispectrum model, however, we expect these effects to be subdominant since we restrict ourselves to the linear regime. We also only considered the squeezed-limit configuration of the bispectrum, which has been shown to dominate the information content of the cross-bispectrum \cite{chiang2015}. Further, we adopt a specific model for the HI bias following \cite{penin}, and the investigation of alternative HI bias models is left for future work. 

In future work, we plan to study constraints on HI-stellar mass scaling relations and their evolution with redshift obtained from the cross-bispectrum \cite{wolz2017determining}. We can also explore how the cross-shot noise term can probe the HI content of optical galaxies \cite{wolz2016intensity, wolz2017determining}. Furthermore, future work will assess how intrinsic alignment systematics will impact the resulting parameter constraints \cite{troxel2015intrinsic}. We will also explore further applications of the cross-bispectrum, for example, we are currently studying constraints on beyond-$\Lambda$CDM parameters such as primordial non-Gaussianity \cite{karagiannis2018constraining, Randrianjanahary2026}, neutrino mass \cite{moodley2023crossbispectrum,kamalinejad2025neutrino}, and modified gravity through structure formation \cite{yamauchi2017constraining, Naidoo2026}, and also aim to examine how the HI-galaxy density cross-bispectrum can be used to improve photometric redshifts calibration \cite{Alonso, guandalin2022clustering}.

\acknowledgments 
WN and MA acknowledges the financial assistance of the South African Radio Astronomy Observatory (SARAO) towards this research (www.sarao.ac.za).
KM acknowledges research funding support from the National Research Foundation, South Africa. WN, MA, and KM thank the Rubin Observatory collaboration for the resources and research interactions with individual members.

\FloatBarrier
\bibliographystyle{JHEP}
\bibliography{galaxy_density_21_cross}

\providecommand{\href}[2]{#2}\begingroup\raggedright\begin{thebibliography}{100}

\bibitem{battye2013h}
R.~Battye, I.~Browne, C.~Dickinson, G.~Heron, B.~Maffei and A.~Pourtsidou,
  \emph{{H I} intensity mapping: a single dish approach}, {\emph{Monthly
  Notices of the Royal Astronomical Society} {\bfseries 434} (2013) 1239}.

\bibitem{Bull}
P.~{Bull}, P.G.~{Ferreira}, P.~{Patel} and M.G.~{Santos}, \emph{{Late-time
  Cosmology with 21 cm Intensity Mapping Experiments}},
  \href{https://doi.org/10.1088/0004-637X/803/1/21}{\emph{The Astrophysical
  Journal} {\bfseries 803} (2015) 21}
  [\href{https://arxiv.org/abs/1405.1452}{{\ttfamily 1405.1452}}].

\bibitem{furlanetto2006cosmology}
S.R.~Furlanetto, S.P.~Oh and F.H.~Briggs, \emph{Cosmology at low frequencies:
  The 21 cm transition and the high-redshift universe}, {\emph{Physics reports}
  {\bfseries 433} (2006) 181}.

\bibitem{crichton2022hydrogen}
D.~Crichton, M.~Aich, A.~Amara, K.~Bandura, B.A.~Bassett, C.~Bengaly et~al.,
  \emph{Hydrogen intensity and real-time analysis experiment: 256-element array
  status and overview}, {\emph{Journal of Astronomical Telescopes, Instruments,
  and Systems} {\bfseries 8} (2022) 011019}.

\bibitem{ska}
D.J.~Bacon, R.A.~Battye, P.~Bull et~al., \emph{Cosmology with phase 1 of the
  square kilometre array red book 2018: Technical specifications and
  performance forecasts},
  \href{https://doi.org/10.1017/pasa.2019.51}{\emph{Publications of the
  Astronomical Society of Australia} {\bfseries 37} (2020) }.

\bibitem{chen2012tianlai}
X.~Chen, \emph{The tianlai project: a 21cm cosmology experiment},  in
  \emph{International Journal of Modern Physics: Conference Series}, vol.~12,
  pp.~256--263, World Scientific, 2012.

\bibitem{battye2016update}
R.~Battye, I.~Browne, T.~Chen, C.~Dickinson, S.~Harper, L.~Olivari et~al.,
  \emph{Update on the bingo 21cm intensity mapping experiment}, {\emph{arXiv
  preprint arXiv:1610.06826} (2016) }.

\bibitem{bandura2014canadian}
K.~Bandura, G.E.~Addison, M.~Amiri, J.R.~Bond, D.~Campbell-Wilson, L.~Connor
  et~al., \emph{Canadian hydrogen intensity mapping experiment (chime)
  pathfinder},  in \emph{Ground-based and Airborne Telescopes V}, vol.~9145,
  pp.~738--757, SPIE, 2014.

\bibitem{vanderlinde2019lrp}
K.~Vanderlinde, K.~Bandura, L.~Belostotski, R.~Bond, P.~Boyle, J.~Brown et~al.,
  \emph{Lrp 2020 whitepaper: The canadian hydrogen observatory and
  radio-transient detector (chord)}, {\emph{arXiv preprint arXiv:1911.01777}
  (2019) }.

\bibitem{NAN_2011}
R.~Nan, D.~Li, C.~Jin, Q.~Wang, L.~Zhu, W.~Zhu et~al., \emph{The
  five-hundred-meter aperture spherical radio telescope ({FAST}) project}, .

\bibitem{Hu_2020}
W.~Hu, X.~Wang, F.~Wu, Y.~Wang, P.~Zhang and X.~Chen, \emph{Forecast for
  {FAST}: from galaxies survey to intensity mapping},
  \href{https://doi.org/10.1093/mnras/staa650}{\emph{Monthly Notices of the
  Royal Astronomical Society} {\bfseries 493} (2020) 5854}.

\bibitem{lsst}
{LSST Science Collaboration}, P.A.~{Abell}, J.~{Allison}, S.F.~{Anderson}
  et~al., \emph{{LSST Science Book, Version 2.0}}, {\emph{ArXiv e-prints}
  (2009) } [\href{https://arxiv.org/abs/0912.0201}{{\ttfamily 0912.0201}}].

\bibitem{desi}
M.~{Levi}, C.~{Bebek}, T.~{Beers}, R.~{Blum} and et.~al. {representing the DESI
  collaboration}, \emph{{The DESI Experiment, a whitepaper for Snowmass 2013}},
  {\emph{ArXiv e-prints} (2013) }
  [\href{https://arxiv.org/abs/1308.0847}{{\ttfamily 1308.0847}}].

\bibitem{euclid}
R.~{Laureijs}, J.~{Amiaux}, S.~{Arduini}, J..~{Augu{\`e}res} et~al.,
  \emph{{Euclid Definition Study Report}}, {\emph{ArXiv e-prints} (2011) }
  [\href{https://arxiv.org/abs/1110.3193}{{\ttfamily 1110.3193}}].

\bibitem{dark2016dark}
D.E.S.~Collaboration:, T.~Abbott, F.~Abdalla, J.~Aleksi{\'c}, S.~Allam,
  A.~Amara et~al., \emph{The dark energy survey: more than dark energy--an
  overview}, {\emph{Monthly Notices of the Royal Astronomical Society}
  {\bfseries 460} (2016) 1270}.

\bibitem{de2013kilo}
J.T.~de~Jong, G.A.~Verdoes~Kleijn, K.H.~Kuijken, E.A.~Valentijn, KiDS and
  A.-W.~Consortiums, \emph{The kilo-degree survey}, {\emph{Experimental
  Astronomy} {\bfseries 35} (2013) 25}.

\bibitem{aihara2018hyper}
H.~Aihara, N.~Arimoto, R.~Armstrong, S.~Arnouts, N.A.~Bahcall, S.~Bickerton
  et~al., \emph{The hyper suprime-cam ssp survey: overview and survey design},
  {\emph{Publications of the Astronomical Society of Japan} {\bfseries 70}
  (2018) S4}.

\bibitem{wfirst}
D.~{Spergel}, N.~{Gehrels}, C.~{Baltay}, D.~{Bennett} and et. al.,
  \emph{{Wide-Field InfrarRed Survey Telescope-Astrophysics Focused Telescope
  Assets WFIRST-AFTA 2015 Report}}, {\emph{ArXiv e-prints} (2015) }
  [\href{https://arxiv.org/abs/1503.03757}{{\ttfamily 1503.03757}}].

\bibitem{Bartelmann_2001}
M.~Bartelmann and P.~Schneider, \emph{Weak gravitational lensing},
  \href{https://doi.org/10.1016/s0370-1573(00)00082-x}{\emph{Physics Reports}
  {\bfseries 340} (2001) 291}.

\bibitem{lsstde}
L.D.E.S.~Collaboration, \emph{Large synoptic survey telescope: Dark energy
  science collaboration},  2012.

\bibitem{Ivanov_2020}
M.M.~Ivanov, M.~Simonovi{\'{c}} and M.~Zaldarriaga, \emph{Cosmological
  parameters from the {BOSS} galaxy power spectrum},
  \href{https://doi.org/10.1088/1475-7516/2020/05/042}{\emph{Journal of
  Cosmology and Astroparticle Physics} {\bfseries 2020} (2020) 042}.

\bibitem{wyithe2007correlation}
J.S.B.~Wyithe and A.~Loeb, \emph{The correlation between the distribution of
  galaxies and 21-cm emission at high redshifts}, {\emph{Monthly Notices of the
  Royal Astronomical Society} {\bfseries 375} (2007) 1034}.

\bibitem{Chang:2010jp}
T.-C.~Chang, U.-L.~Pen, K.~Bandura and J.B.~Peterson, \emph{{Hydrogen 21-cm
  Intensity Mapping at redshift 0.8}},
  \href{https://doi.org/10.1038/nature09187}{\emph{Nature} {\bfseries 466}
  (2010) 463} [\href{https://arxiv.org/abs/1007.3709}{{\ttfamily 1007.3709}}].

\bibitem{masui2013measurement}
K.~Masui, E.~Switzer, N.~Banavar, K.~Bandura, C.~Blake, L.-M.~Calin et~al.,
  \emph{Measurement of 21 cm brightness fluctuations at z~ 0.8 in
  cross-correlation}, {\emph{The Astrophysical Journal Letters} {\bfseries 763}
  (2013) L20}.

\bibitem{wolz2017determining}
L.~Wolz, C.~Blake and J.~Wyithe, \emph{Determining the {HI} content of galaxies
  via intensity mapping cross-correlations}, {\emph{Monthly Notices of the
  Royal Astronomical Society} {\bfseries 470} (2017) 3220}.

\bibitem{Pourtsidou}
A.~{Pourtsidou}, D.~{Bacon} and R.~{Crittenden}, \emph{{{H I} and cosmological
  constraints from intensity mapping, optical and CMB surveys}},
  \href{https://doi.org/10.1093/mnras/stx1479}{\emph{\mnras} {\bfseries 470}
  (2017) 4251} [\href{https://arxiv.org/abs/1610.04189}{{\ttfamily
  1610.04189}}].

\bibitem{Alonso}
D.~{Alonso}, P.G.~{Ferreira}, M.J.~{Jarvis} and K.~{Moodley},
  \emph{{Calibrating photometric redshifts with intensity mapping
  observations}},
  \href{https://doi.org/10.1103/PhysRevD.96.043515}{\emph{Physical Review D}
  {\bfseries 96} (2017) 043515}
  [\href{https://arxiv.org/abs/1704.01941}{{\ttfamily 1704.01941}}].

\bibitem{padmanabhan2020cross}
H.~Padmanabhan, A.~Refregier and A.~Amara, \emph{Cross-correlating 21 cm and
  galaxy surveys: implications for cosmology and astrophysics}, {\emph{Monthly
  Notices of the Royal Astronomical Society} {\bfseries 495} (2020) 3935}.

\bibitem{fonseca2018synergies}
J.~Fonseca, R.~Maartens and M.G.~Santos, \emph{Synergies between intensity maps
  of hydrogen lines}, {\emph{Monthly Notices of the Royal Astronomical Society}
  {\bfseries 479} (2018) 3490}.

\bibitem{guandalin2022clustering}
C.~Guandalin, I.P.~Carucci, D.~Alonso and K.~Moodley, \emph{Clustering
  redshifts with the 21cm-galaxy cross-bispectrum}, {\emph{Monthly Notices of
  the Royal Astronomical Society} {\bfseries 516} (2022) 3029}.

\bibitem{jalilvand2020new}
M.~Jalilvand, E.~Majerotto, C.~Bonvin, F.~Lacasa, M.~Kunz, W.~Naidoo et~al.,
  \emph{New estimator for gravitational lensing using galaxy and intensity
  mapping surveys}, {\emph{Physical review letters} {\bfseries 124} (2020)
  031101}.

\bibitem{pourtsidou2017h}
A.~Pourtsidou, D.~Bacon and R.~Crittenden, \emph{{HI} and cosmological
  constraints from intensity mapping, optical and cmb surveys}, {\emph{Monthly
  Notices of the Royal Astronomical Society} {\bfseries 470} (2017) 4251}.

\bibitem{bacon2020cosmology}
D.J.~Bacon, R.A.~Battye, P.~Bull, S.~Camera, P.G.~Ferreira, I.~Harrison et~al.,
  \emph{Cosmology with phase 1 of the square kilometre array red book 2018:
  Technical specifications and performance forecasts}, {\emph{Publications of
  the Astronomical Society of Australia} {\bfseries 37} (2020) e007}.

\bibitem{wolz2016intensity}
L.~Wolz, C.~Tonini, C.~Blake and J.~Wyithe, \emph{Intensity mapping
  cross-correlations: connecting the largest scales to galaxy evolution},
  {\emph{Monthly Notices of the Royal Astronomical Society} {\bfseries 458}
  (2016) 3399}.

\bibitem{guo2017constraining}
H.~Guo, C.~Li, Z.~Zheng, H.~Mo, Y.~Jing, Y.~Zu et~al., \emph{Constraining the
  {H I--Halo} mass relation from galaxy clustering}, {\emph{The Astrophysical
  Journal} {\bfseries 846} (2017) 61}.

\bibitem{Pourtsidou2016}
A.~Pourtsidou, D.~Bacon, R.~Crittenden and R.B.~Metcalf, \emph{{Prospects for
  clustering and lensing measurements with forthcoming intensity mapping and
  optical surveys}}, \href{https://doi.org/10.1093/mnras/stw658}{\emph{Monthly
  Notices of the Royal Astronomical Society} {\bfseries 459} (2016) 863}.

\bibitem{Pourtsidou2015}
A.~Pourtsidou, D.~Bacon and R.~Crittenden, \emph{Cross-correlation cosmography
  with intensity mapping of the neutral hydrogen 21 cm emission},
  \href{https://doi.org/10.1103/PhysRevD.92.103506}{\emph{Phys. Rev. D}
  {\bfseries 92} (2015) 103506}.

\bibitem{ansari2018inflation}
R.~Ansari, E.J.~Arena, K.~Bandura, P.~Bull, E.~Castorina, T.-C.~Chang et~al.,
  \emph{Inflation and early dark energy with a stage ii hydrogen intensity
  mapping experiment}, {\emph{arXiv preprint arXiv:1810.09572} (2018) }.

\bibitem{shi2020hir4}
F.~Shi, Y.-S.~Song, J.~Asorey, D.~Parkinson, K.~Ahn, J.~Yao et~al., \emph{Hir4:
  cosmological signatures imprinted on the cross-correlation between a 21-cm
  map and galaxy clustering}, {\emph{Monthly Notices of the Royal Astronomical
  Society} {\bfseries 499} (2020) 4613}.

\bibitem{viljoen2021multi}
J.-A.~Viljoen, J.~Fonseca and R.~Maartens, \emph{Multi-wavelength spectroscopic
  probes: prospects for primordial non-gaussianity and relativistic effects},
  {\emph{Journal of Cosmology and Astroparticle Physics} {\bfseries 2021}
  (2021) 010}.

\bibitem{fang2022cosmology}
X.~Fang, T.~Eifler, E.~Schaan, H.-J.~Huang, E.~Krause and S.~Ferraro,
  \emph{Cosmology from clustering, cosmic shear, cmb lensing, and cross
  correlations: combining rubin observatory and simons observatory},
  {\emph{Monthly Notices of the Royal Astronomical Society} {\bfseries 509}
  (2022) 5721}.

\bibitem{sgier2021combined}
R.~Sgier, C.~Lorenz, A.~Refregier, J.~Fluri, D.~Z{\"u}rcher and F.~Tarsitano,
  \emph{Combined $13\times2$-point analysis of the cosmic microwave background
  and large-scale structure: implications for the $s_8$-tension and neutrino
  mass constraints}, {\emph{arXiv preprint arXiv:2110.03815} (2021) }.

\bibitem{white2022cosmological}
M.~White, R.~Zhou, J.~DeRose, S.~Ferraro, S.-F.~Chen, N.~Kokron et~al.,
  \emph{Cosmological constraints from the tomographic cross-correlation of desi
  luminous red galaxies and planck cmb lensing}, {\emph{Journal of Cosmology
  and Astroparticle Physics} {\bfseries 2022} (2022) 007}.

\bibitem{Santos_2005}
M.G.~Santos, A.~Cooray and L.~Knox, \emph{Multifrequency analysis of 21
  centimeter fluctuations from the era of reionization},
  \href{https://doi.org/10.1086/429857}{\emph{The Astrophysical Journal}
  {\bfseries 625} (2005) 575}.

\bibitem{Shaw_2014}
J.R.~Shaw, K.~Sigurdson, U.-L.~Pen, A.~Stebbins and M.~Sitwell,
  \emph{{All}-{Sky} {Interferometry} {With} {Spherical} {Harmonic} {Transit}
  {Telescopes}}, \href{https://doi.org/10.1088/0004-637x/781/2/57}{\emph{The
  Astrophysical Journal} {\bfseries 781} (2014) 57}.

\bibitem{Liu2013}
A.~Liu, J.R.~Pritchard, M.~Tegmark and A.~Loeb, \emph{Global 21 cm signal
  experiments: A designer's guide},
  \href{https://doi.org/10.1103/PhysRevD.87.043002}{\emph{Phys. Rev. D}
  {\bfseries 87} (2013) 043002}.

\bibitem{Shaw_2015}
J.R.~Shaw, K.~Sigurdson, M.~Sitwell, A.~Stebbins and U.-L.~Pen, \emph{Coaxing
  cosmic 21~cm fluctuations from the polarized sky using $m$-mode analysis},
  \href{https://doi.org/10.1103/physrevd.91.083514}{\emph{Physical Review D}
  {\bfseries 91} (2015) }.

\bibitem{Switzer_2014}
E.R.~Switzer and A.~Liu, \emph{Erasing the variable: Empirical foreground
  discovery for global 21 cm spectrum experiments},
  \href{https://doi.org/10.1088/0004-637X/793/2/102}{\emph{The Astrophysical
  Journal} {\bfseries 793} (2014) 102}.

\bibitem{Parsons_2012}
A.R.~Parsons, J.C.~Pober, J.E.~Aguirre, C.L.~Carilli, D.C.~Jacobs and
  D.F.~Moore, \emph{A per-baseline, delay-spectrum technique for accessing the
  21 cm cosmic reionization signature},
  \href{https://doi.org/10.1088/0004-637X/756/2/165}{\emph{The Astrophysical
  Journal} {\bfseries 756} (2012) 165}.

\bibitem{moodley2023crossbispectrum}
K.~Moodley, W.~Naidoo, H.~Prince and A.~Penin, \emph{A cross-bispectrum
  estimator for cmb-hi intensity mapping correlations},  2023.

\bibitem{kothari2024wide}
R.~Kothari and R.~Maartens, \emph{A wide-angle formulation of foreground
  filters for hi intensity mapping}, {\emph{Journal of Cosmology and
  Astroparticle Physics} {\bfseries 2024} (2024) 089}.

\bibitem{shen2026direct}
D.~Shen, N.~Kokron and E.~Schaan, \emph{Direct correlation of line intensity
  mapping and cmb lensing from evolution along the lightcone}, {\emph{Physical
  Review D} {\bfseries 113} (2026) 023521}.

\bibitem{anderson2018low}
C.~Anderson, N.~Luciw, Y.-C.~Li, C.~Kuo, J.~Yadav, K.~Masui et~al.,
  \emph{Low-amplitude clustering in low-redshift 21-cm intensity maps
  cross-correlated with 2df galaxy densities}, {\emph{Monthly Notices of the
  Royal Astronomical Society} {\bfseries 476} (2018) 3382}.

\bibitem{Amiri_2023}
T.C.~Collaboration, M.~Amiri, K.~Bandura, T.~Chen, M.~Deng, M.~Dobbs et~al.,
  \emph{Detection of cosmological 21 cm emission with the canadian hydrogen
  intensity mapping experiment},
  \href{https://doi.org/10.3847/1538-4357/acb13f}{\emph{The Astrophysical
  Journal} {\bfseries 947} (2023) 16}.

\bibitem{cunnington2023h}
S.~Cunnington, Y.~Li, M.G.~Santos, J.~Wang, I.P.~Carucci, M.O.~Irfan et~al.,
  \emph{{HI} intensity mapping with meerkat: power spectrum detection in
  cross-correlation with wigglez galaxies}, {\emph{Monthly Notices of the Royal
  Astronomical Society} {\bfseries 518} (2023) 6262}.

\bibitem{paul2026direct}
S.~Paul, Z.~Chen, M.G.~Santos and L.~Wolz, \emph{A direct detection of neutral
  hydrogen intensity mapping on mpc scales at z$\approx$ 0.32 and z$\approx$
  0.44}, {\emph{The Astrophysical Journal Letters} {\bfseries 1005} (2026)
  L56}.

\bibitem{Bernardeau_2002}
F.~Bernardeau, S.~Colombi, E.~Gazta{\~{n}}aga and R.~Scoccimarro,
  \emph{Large-scale structure of the universe and cosmological perturbation
  theory}, \href{https://doi.org/10.1016/s0370-1573(02)00135-7}{\emph{Physics
  Reports} {\bfseries 367} (2002) 1}.

\bibitem{chiang2014}
C.-T.~Chiang, C.~Wagner, F.~Schmidt and E.~Komatsu, \emph{Position-dependent
  power spectrum of the large-scale structure: a novel method to measure the
  squeezed-limit bispectrum},
  \href{https://doi.org/10.1088/1475-7516/2014/05/048}{\emph{Journal of
  Cosmology and Astroparticle Physics} {\bfseries 2014} (2014) 048}.

\bibitem{chiang2015}
C.-T.~Chiang, \emph{Position-dependent power spectrum: a new observable in the
  large-scale structure},  2015.

\bibitem{Takada2013}
M.~Takada and W.~Hu, \emph{Power spectrum super-sample covariance},
  \href{https://doi.org/10.1103/PhysRevD.87.123504}{\emph{Phys. Rev. D}
  {\bfseries 87} (2013) 123504}.

\bibitem{Takada_2003}
M.~Takada and B.~Jain, \emph{Three-point correlations in weak lensing surveys:
  model predictions and applications},
  \href{https://doi.org/10.1046/j.1365-8711.2003.06868.x}{\emph{Monthly Notices
  of the Royal Astronomical Society} {\bfseries 344} (2003) 857}.

\bibitem{Chakraborty_2026}
T.C.~Collaboration, A.~Chakraborty, M.~Dobbs, S.~Foreman, L.~Gray, M.~Halpern
  et~al., \emph{The squeezed bispectrum from chime {H I}f emission and planck
  cosmic microwave background lensing: Current sensitivity and forecasts},
  \href{https://doi.org/10.3847/1538-4357/ae77e6}{\emph{The Astrophysical
  Journal} {\bfseries 1005} (2026) 73}.

\bibitem{Aghanim}
{Planck Collaboration}, {Aghanim, N.}, {Akrami, Y.}, {Ashdown, M.}, {Aumont,
  J.} et~al., \emph{Planck 2018 results - vi. cosmological parameters},
  \href{https://doi.org/10.1051/0004-6361/201833910}{\emph{A\&A} {\bfseries
  641} (2020) A6}.

\bibitem{penin}
A.~P{\'e}nin, O.~Umeh and M.G.~Santos, \emph{{A scale-dependent bias on linear
  scales: the case for H I intensity mapping at z = 1}},
  \href{https://doi.org/10.1093/mnras/stx2635}{\emph{Monthly Notices of the
  Royal Astronomical Society} {\bfseries 473} (2017) 4297}.

\bibitem{Hogg}
D.W.~{Hogg}, \emph{{Distance measures in cosmology}}, {\emph{ArXiv Astrophysics
  e-prints} (1999) } [\href{https://arxiv.org/abs/astro-ph/9905116}{{\ttfamily
  astro-ph/9905116}}].

\bibitem{Tegmark_2002}
M.~Tegmark, S.~Dodelson, D.J.~Eisenstein and et~al. (for~the
  SDSS~Collaboration), \emph{The angular power spectrum of galaxies from early
  sloan digital sky survey data},
  \href{https://doi.org/10.1086/339894}{\emph{The Astrophysical Journal}
  {\bfseries 571} (2002) 191}.

\bibitem{Peiris_2000}
H.V.~Peiris and D.N.~Spergel, \emph{Cross-correlating the sloan digital sky
  survey with the microwave sky},
  \href{https://doi.org/10.1086/309373}{\emph{The Astrophysical Journal}
  {\bfseries 540} (2000) 605}.

\bibitem{Kaiser_1998}
N.~Kaiser, \emph{Weak lensing and cosmology},
  \href{https://doi.org/10.1086/305515}{\emph{The Astrophysical Journal}
  {\bfseries 498} (1998) 26}.

\bibitem{Barber}
A.J.~Barber and A.N.~Taylor, \emph{{Shear and magnification angular power
  spectra and higher-order moments from weak gravitational lensing}},
  \href{https://doi.org/10.1046/j.1365-8711.2003.06872.x}{\emph{Monthly Notices
  of the Royal Astronomical Society} {\bfseries 344} (2003) 789}.

\bibitem{Joachimi_2010}
B.~Joachimi and S.L.~Bridle, \emph{Simultaneous measurement of cosmology and
  intrinsic alignments using joint cosmic shear and galaxy number density
  correlations},
  \href{https://doi.org/10.1051/0004-6361/200913657}{\emph{Astronomy \&amp;
  Astrophysics} {\bfseries 523} (2010) A1}.

\bibitem{Kilbinger_2015}
M.~Kilbinger, \emph{Cosmology with cosmic shear observations: a review},
  \href{https://doi.org/10.1088/0034-4885/78/8/086901}{\emph{Reports on
  Progress in Physics} {\bfseries 78} (2015) 086901}.

\bibitem{Kilbinger_2017}
M.~Kilbinger, C.~Heymans, M.~Asgari, S.~Joudaki, P.~Schneider, P.~Simon et~al.,
  \emph{Precision calculations of the cosmic shear power spectrum projection},
  \href{https://doi.org/10.1093/mnras/stx2082}{\emph{Monthly Notices of the
  Royal Astronomical Society} {\bfseries 472} (2017) 2126}.

\bibitem{fang}
W.~Fang and Z.~Haiman, \emph{Constraining dark energy by combining cluster
  counts and shear-shear correlations in a weak lensing survey},
  \href{https://doi.org/10.1103/PhysRevD.75.043010}{\emph{Phys. Rev. D}
  {\bfseries 75} (2007) 043010}.

\bibitem{song}
Y.-S.~{Song} and L.~{Knox}, \emph{{Determination of cosmological parameters
  from cosmic shear data}},
  \href{https://doi.org/10.1103/PhysRevD.70.063510}{\emph{Physical Review D}
  {\bfseries 70} (2004) 063510}
  [\href{https://arxiv.org/abs/astro-ph/0312175}{{\ttfamily
  astro-ph/0312175}}].

\bibitem{Sherwin_2012}
B.D.~Sherwin, S.~Das, A.~Hajian, G.~Addison, J.R.~Bond, D.~Crichton et~al.,
  \emph{The atacama cosmology telescope: Cross-correlation of cosmic microwave
  background lensing and quasars},
  \href{https://doi.org/10.1103/physrevd.86.083006}{\emph{Physical Review D}
  {\bfseries 86} (2012) }.

\bibitem{mandelbaum2018lsst}
R.~Mandelbaum, T.~Eifler, R.~Hlo{\v{z}}ek, T.~Collett, E.~Gawiser, D.~Scolnic
  et~al., \emph{The lsst dark energy science collaboration (desc) science
  requirements document}, {\emph{arXiv preprint arXiv:1809.01669} (2018) }.

\bibitem{Kaiser_1992}
N.~{Kaiser}, \emph{{Weak Gravitational Lensing of Distant Galaxies}},
  \href{https://doi.org/10.1086/171151}{\emph{The Astrophysical Journal}
  {\bfseries 388} (1992) 272}.

\bibitem{Hu_2001}
W.~Hu, \emph{Dark synergy: Gravitational lensing and the cmb},
  \href{https://doi.org/10.1103/physrevd.65.023003}{\emph{Physical Review D}
  {\bfseries 65} (2001) }.

\bibitem{tegmark1997measuring}
M.~Tegmark, \emph{Measuring cosmological parameters with galaxy surveys},
  {\emph{Physical Review Letters} {\bfseries 79} (1997) 3806}.

\bibitem{biagetti2022covariance}
M.~Biagetti, L.~Castiblanco, J.~Nore{\~n}a and E.~Sefusatti, \emph{The
  covariance of squeezed bispectrum configurations}, {\emph{Journal of
  Cosmology and Astroparticle Physics} {\bfseries 2022} (2022) 009}.

\bibitem{holwerda2011looking}
B.~Holwerda, S.-L.~Blyth, A.~Baker et~al., \emph{Looking at the distant
  universe with the meerkat array (laduma)}, {\emph{Proceedings of the
  International Astronomical Union} {\bfseries 7} (2011) 496}.

\bibitem{ponomareva2023mightee}
A.A.~Ponomareva, M.J.~Jarvis, H.~Pan, N.~Maddox, M.G.~Jones, B.S.~Frank et~al.,
  \emph{{MIGHTEE-H I}: the first meerkat {H I} mass function from an untargeted
  interferometric survey}, {\emph{Monthly Notices of the Royal Astronomical
  Society} {\bfseries 522} (2023) 5308}.

\bibitem{blake2003probing}
C.~Blake and K.~Glazebrook, \emph{Probing dark energy using baryonic
  oscillations in the galaxy power spectrum as a cosmological ruler},
  {\emph{The Astrophysical Journal} {\bfseries 594} (2003) 665}.

\bibitem{sarkar2019modelling}
D.~Sarkar, S.~Majumdar and S.~Bharadwaj, \emph{Modelling the post-reionization
  neutral hydrogen ({HI}) 21-cm bispectrum}, {\emph{Monthly Notices of the
  Royal Astronomical Society} {\bfseries 490} (2019) 2880}.

\bibitem{villaescusa2018ingredients}
F.~Villaescusa-Navarro, S.~Genel, E.~Castorina, A.~Obuljen, D.N.~Spergel,
  L.~Hernquist et~al., \emph{Ingredients for 21 cm intensity mapping},
  {\emph{The Astrophysical Journal} {\bfseries 866} (2018) 135}.

\bibitem{zhou2024parametrization}
S.~Zhou and P.~Zhang, \emph{Parametrization of stochasticity in galaxy
  clustering and reconstruction of tomographic matter clustering},
  {\emph{Physical Review D} {\bfseries 110} (2024) 123528}.

\bibitem{alcock1979evolution}
C.~Alcock and B.~Paczy{\'n}ski, \emph{An evolution free test for non-zero
  cosmological constant}, {\emph{Nature} {\bfseries 281} (1979) 358}.

\bibitem{abbott2018dark}
T.~Abbott, F.~Abdalla, A.~Alarcon et~al., \emph{Dark energy survey year 1
  results: Joint analysis of galaxy clustering, galaxy lensing, and cmb lensing
  two-point functions}, {\emph{arXiv preprint arXiv:1810.02322} (2018) }.

\bibitem{descollaboration2026darkenergysurveyyear}
D.~Collaboration, T.M.C.~Abbott, M.~Adamow, M.~Aguena, A.~Alarcon, S.S.~Allam
  et~al., \emph{Dark energy survey year 6 results: Cosmological constraints
  from galaxy clustering and weak lensing},  2026.

\bibitem{albrecht2006report}
A.~Albrecht, G.~Bernstein, R.~Cahn et~al., \emph{Report of the dark energy task
  force},  2006.

\bibitem{de2017vimos}
S.~De~La~Torre, E.~Jullo, C.~Giocoli, A.~Pezzotta, J.~Bel, B.~Granett et~al.,
  \emph{The vimos public extragalactic redshift survey (vipers)-gravity test
  from the combination of redshift-space distortions and galaxy-galaxy lensing
  at 0.5< z< 1.2}, {\emph{Astronomy \& Astrophysics} {\bfseries 608} (2017)
  A44}.

\bibitem{jullo2019testing}
E.~Jullo, S.~De~La~Torre, M.-C.~Cousinou, S.~Escoffier, C.~Giocoli,
  R.B.~Metcalf et~al., \emph{Testing gravity with galaxy-galaxy lensing and
  redshift-space distortions using cfht-stripe 82, cfhtlens, and boss cmass
  datasets}, {\emph{Astronomy \& Astrophysics} {\bfseries 627} (2019) A137}.

\bibitem{gil2016clustering}
H.~Gil-Mar{\'\i}n, W.J.~Percival, L.~Verde, J.R.~Brownstein, C.-H.~Chuang,
  F.-S.~Kitaura et~al., \emph{The clustering of galaxies in the sdss-iii baryon
  oscillation spectroscopic survey: Rsd measurement from the power spectrum and
  bispectrum of the dr12 boss galaxies}, {\emph{Monthly Notices of the Royal
  Astronomical Society} (2016) stw2679}.

\bibitem{byun2020constraining}
J.~Byun, F.O.~Franco, C.~Howlett, C.~Bonvin and D.~Obreschkow,
  \emph{Constraining the growth rate of structure with phase correlations},
  {\emph{Monthly Notices of the Royal Astronomical Society} {\bfseries 497}
  (2020) 1765}.

\bibitem{perenon2020improvements}
L.~Perenon, S.~Ili{\'c}, R.~Maartens and A.~de~La~Cruz-Dombriz,
  \emph{Improvements in cosmological constraints from breaking growth
  degeneracy}, {\emph{Astronomy \& Astrophysics} {\bfseries 642} (2020) A116}.

\bibitem{adame2025desi}
A.~Adame, J.~Aguilar, S.~Ahlen, S.~Alam, D.~Alexander, M.~Alvarez et~al.,
  \emph{Desi 2024 v: Full-shape galaxy clustering from galaxies and quasars},
  {\emph{Journal of Cosmology and Astroparticle Physics} {\bfseries 2025}
  (2025) 008}.

\bibitem{wolz2022h}
L.~Wolz, A.~Pourtsidou, K.W.~Masui, T.-C.~Chang, J.E.~Bautista, E.-M.~Mueller
  et~al., \emph{{HI} constraints from the cross-correlation of eboss galaxies
  and green bank telescope intensity maps}, {\emph{Monthly Notices of the Royal
  Astronomical Society} {\bfseries 510} (2022) 3495}.

\bibitem{alam2021completed}
S.~Alam, M.~Aubert, S.~Avila, C.~Balland, J.E.~Bautista, M.A.~Bershady et~al.,
  \emph{Completed sdss-iv extended baryon oscillation spectroscopic survey:
  Cosmological implications from two decades of spectroscopic surveys at the
  apache point observatory}, {\emph{Physical Review D} {\bfseries 103} (2021)
  083533}.

\bibitem{vargas2019unraveling}
M.~Vargas-Magana, D.D.~Brooks, M.M.~Levi and G.G.~Tarle, \emph{Unraveling the
  universe with desi}, {\emph{arXiv preprint arXiv:1901.01581} (2019) }.

\bibitem{ilic2022euclid}
S.~Ili{\'c}, N.~Aghanim, C.~Baccigalupi, J.R.~Bermejo-Climent, G.~Fabbian,
  L.~Legrand et~al., \emph{Euclid preparation-xv. forecasting cosmological
  constraints for the euclid and cmb joint analysis}, {\emph{Astronomy \&
  Astrophysics} {\bfseries 657} (2022) A91}.

\bibitem{troxel2015intrinsic}
M.~Troxel and M.~Ishak, \emph{The intrinsic alignment of galaxies and its
  impact on weak gravitational lensing in an era of precision cosmology},
  {\emph{Physics Reports} {\bfseries 558} (2015) 1}.

\bibitem{karagiannis2018constraining}
D.~Karagiannis, A.~Lazanu, M.~Liguori, A.~Raccanelli, N.~Bartolo and L.~Verde,
  \emph{Constraining primordial non-gaussianity with bispectrum and power
  spectrum from upcoming optical and radio surveys}, {\emph{Monthly Notices of
  the Royal Astronomical Society} {\bfseries 478} (2018) 1341}.

\bibitem{Randrianjanahary2026}
L.~Randrianjanahary, W.~Naidoo and et~al., ``{Probing $f_{\rm NL}$ using
  HI-HI-Shear Integrated cross-bispectrum with HIRAX and LSST}.'' 2026.

\bibitem{kamalinejad2025neutrino}
F.~Kamalinejad and Z.~Slepian, \emph{Neutrino mass signatures in the galaxy
  bispectrum}, {\emph{Physical Review D} {\bfseries 112} (2025) 083501}.

\bibitem{yamauchi2017constraining}
D.~Yamauchi, S.~Yokoyama and H.~Tashiro, \emph{Constraining modified theories
  of gravity with the galaxy bispectrum}, {\emph{Physical Review D} {\bfseries
  96} (2017) 123516}.

\bibitem{Naidoo2026}
W.~Naidoo and et~al., ``{Bispectral correlations between HI intensity mapping
  and CMB lensing}.'' 2026.

\end{thebibliography}\endgroup
\end{document}